\documentclass[epj,nopacs]{svjour}
\usepackage{epsfig,array}
\def\nn{\nonumber \\}

\def\fun#1#2{\lower3.6pt\vbox{\baselineskip0pt\lineskip.9pt
\ialign{$\mathsurround=0pt#1\hfil##\hfil$\crcr#2\crcr\sim\crcr}}}

\begin{document}
\newcommand{\be}{\begin{eqnarray}}
\newcommand{\ee}{\end{eqnarray}}
\newcommand{\inli}{\int\limits}

%\begin{frontmatter}

\date{\today}

\title{\bf
Partial decay widths of baryons in the spin-momentum operator expansion method}

\author
{
A.V.~Anisovich \inst{1,2}
\and A.V.~Sarantsev \inst{1,2}
}

\institute
{
HISKP, Universit\"at Bonn, D-53115 Bonn,
\and Petersburg Nuclear Physics Institute, Gatchina, Russia
}

%-------------abstract----------------

\abstract{The cross sections for photo- and pion-induced production of
baryon resonances and their partial decay widths to the two--body and
multi--body final states are calculated in the framework of the
operator expansion method. The approach is fully relativistic
invariant, and it allows us to perform combined analyses of different
reactions imposing directly the analyticity and unitarity constraints. All
formulae are given explicitly in the form used by the Crystal Barrel
collaboration in the partial wave analysis.} %\vspace{.5cm}

%\it{PACS: 13.75.-n, 25.43.+t }

%----------end of abstract-------------
\titlerunning{\bf
Partial decay widths of baryons in the spin-momentum operator...}

\mail{andsar@hiskp.uni-bonn.de}

%\begin{document}
\maketitle

\section{Introduction}

The detailed knowledge of the low-energy hadron masses
and their decays is vital to construct and test QCD-inspired models
for the nonperturbative regime. In recent years, appreciable progress
has been achieved in meson spectroscopy where
a new information was
mostly obtained from the analysis of reactions with three or more
particles in the final state. A prime example is the rich spectrum
of new resonances obtained in the $\bar pp$ annihilation at rest
\cite{Amsler:1994rv,Anisovich:1994bi,Amsler:1994pz,Amsler:1994ah,Amsler:1995bf,Amsler:1995gf,Amsler:1995bz}
and in flight
\cite{Anisovich:2000ut,Anisovich:2001pn,Anisovich:2002su,Anisovich:2002sv}.

While in the meson sector the majority of states
predicted by the quark models
\cite{Godfrey:1985xj,Ricken:2000kf}
has been observed as well as some extra
states with respect to the $q\bar q$ classification,
the situation in the baryon sector is
dramatically different. Here a number of quark models predicts
\cite{Isgur:1995ei,Capstick:bm,Loring:2001ky} much richer spectrum
than that observed so far. In addition, certain models predict the
existence of states which are -- from the quark model point of view --
exotic, like pentaquarks \cite{Diakonov:1997mm},
heptaquarks \cite{Bicudo:2003rw},
or nucleon--meson bound states \cite{Jido:2003cb}.
The lack of the predicted states can be an indication of new
phenomena such as formation of diquark states
\cite{AGS,Klempt:2002vp,book} or specific interaction of colour
particles at large distances
\cite{Klempt:2002vp,Anisovich:2000kx} that lead to linear
trajectories on $(n,M^2)$ (where $n$ is the number of a
radial excitation) and  $(J,M^2)$ planes.
However, the explanation could be much simpler: the modern
knowledge of the baryon spectrum is based mostly on the analysis of
pion-induced reactions on the single meson production (mostly data
obtained from the elastic pion-nucleon scattering) and it is quite
possible that many states with weak coupling to the $\pi N$ channel
escaped the detection.

Recently, a number of experiments on baryon
spectroscopy has been initiated in several laboratories. A large
amount of new data is coming from photoproduction
experiments such as CLAS, CB--ELSA, GRAAL, Mainz--TAPS, LEPS
(see e.g.
\cite{Bartholomy:04,Crede:04,Krusche:nv,Beck,GRAAL1,Rebreyend,GRAAL2,Glander:2003jw,Lawall:2005np,McNabb:2003nf,carnahan,Zegers:2003ux}
). The analysis of the new photoproduction data and especially the data
obtained from reactions with multiparticle final states is therefore
one of the most urgent tasks in baryon spectroscopy.

The meson spectroscopy teaches us that the analysis of
reactions with multiparticle final states cannot be done unambiguously
without information about reactions with the two-body final states. The
best way to impose such information is to perform a combined analysis
of the set of reactions. This issue is even more important in
baryon spectroscopy where often the polarization of initial or/and final
particles is not detected. Here, even in the analysis of the two-body
final states the combined analysis of the data from different channels
plays a vital role. Thus, the development of a method which describes
different reactions on the same basis is a key point in the search for
new baryon states.

The method based on relativistic invariant operators
which are constructed directly from the 4-vectors of the particles
was put forward in a set of articles (see \cite{operators} and
reference therein). In the article \cite{operators}, the
operators for the photoproduction and pion-induced reactions were
introduced and the amplitude angular
dependences were calculated. The analysis of a large number of
single meson photoproduction reactions performed in the framework
of this method \cite{Anisovich:2005tf,Sarantsev:2005tg}
reveals a number of new baryon
states. Here, the resonances were parameterized as T-matrix poles
with mass, width and the product of initial and final couplings as
fitted parameters.
However, such an observation should be confirmed by
the combined analysis of the photoproduction and pion-induced reaction,
with two and many particle final states.
In this case, the couplings of the baryon resonances to different
channels can be unambiguously defined, partial widths
 calculated  and checked for consistency.

In this article, we develop the approach of \cite{operators}
providing explicit expressions for the calculation of cross
sections for different reactions and partial widths of the states.
The obtained expressions can be directly used in the T- or K-matrix
combined analysis of the photo and pion induced reactions.

The article is organized as follows. In section 2, we provide
general expressions for the calculation of cross sections and
resonance widths in the framework of the operator approach. In
section 3, we discuss  resonance widths calculated with
dispersion relations and give explicit expressions for the
resonances decaying into spinless particles. The calculation of $\pi N$
partial widths of baryon states is given in Section 4 and $\gamma N$
partial widths in Section 5. The three-particle partial widths
and correspondent singularities are
discussed in Section 6.

\section{Cross section for transition amplitudes and the
widths of resonances}

For the production of $m$ particles with the 4-momenta
$q_i$ from two particles colliding with
4-momenta $k_1$ and $k_2$, the cross section  is given by
\be
d\sigma=\frac{(2\pi)^4|A|^2}{4|\vec k|\sqrt{s}}\,
d\Phi_m(P,q_1\ldots q_m)\;, \qquad P\!=\!k_1\!+\!k_2\;,
\ee
where $A$ is the transition amplitude, $\vec k$ is the 3-momentum of
the initial particle calculated in the centre--of--mass system (c.m.s.)
of the reaction, and $s=P^2=(k_1+k_2)^2$. The cross section is
calculated by averaging over polarizations of initial--state particles
and summing over polarization of final--state particles. The invariant
m-particle phase space is given by
\be
d\Phi_m(P,q_1\ldots q_m)\! =\! \delta^4(P- \sum\limits_{i=1}^m q_i)\!
\prod\limits_{i=1}^m \! \frac{d^3q_i}{(2\pi)^3 2q_{0i}}\;. \label{phase}
\ee
The amplitude for the transition from the initial state $"in"$ to the
final state $"out"$ via a resonance with the total spin $J$, mass $M$
and width $\Gamma_{tot}$ has the form:
\be
A=\frac{g_{in}Q^{in}_{\mu_1\ldots\mu_n}
F^{\mu_1\ldots\mu_n}_{\nu_1\ldots\nu_n}
Q^{out}_{\nu_1\ldots\nu_n}g_{out}}
{M^2-s-iM\Gamma_{tot}}\,.
\label{tr_amp}
\ee
Here $n=J$ for mesons,  and $n=J-1/2$ for baryons, $g_{in}$ and
$g_{out}$ are initial-- and final--state couplings, $Q^{in}$ and
$Q^{out}$ are operators which describe the production and decay
processes and $F^{\mu_1\ldots\mu_n}_{\nu_1\ldots\nu_n}$ is the tensor
part of the resonance propagator.

The standard formula for the decay of a resonance into $m$ particles
is given by
\be
M\Gamma=\int\!\frac{(2\pi)^4}{2}\;|A_{dec}|^2
d\Phi_m(P,q_1\ldots q_m)
\label{width_def}
\ee
and, as in the case of the cross section, one has to sum over the
polarizations of the final--state particles.

In the operator representation, the amplitude $A_{dec}$ has the form
\be
A_{dec}=
\bar \Psi^{(i)}_{\mu_1 \ldots\mu_n}
Q_{\mu_1\ldots\mu_n}g.
\ee
where $\Psi^{(i)}_{\mu_1 \ldots\mu_n}$
is polarization tensor of the resonance (conventionally, we call it the polarization wave
function), $Q_{\mu_1\ldots\mu_n}$ is the operator of the transition of the resonance into
final state, and
 $g$ is the
corresponding coupling constant. For example, if $Q=Q^{out}$ and
$g=g_{out}$, eq. (\ref{width_def}) provides the partial width for
the resonance decay into the final state
and for $Q=Q^{in}$ and $g=g_{in}$ the partial widths
for its decay into the initial state.

The tensor part of the propagator is determined by the polarization
tensor as follows:
\be
F^{\mu_1\ldots\mu_n}_{\nu_1\ldots\nu_n}=\sum\limits_{i=1}^{2J+1}
\Psi^{(i)}_{\mu_1 \ldots\mu_n} \bar \Psi^{(i)}_{\nu_1 \ldots\nu_n}\;,
\label{psi_psi}
\ee
where the summation is performed over all possible polarizations of the
resonance. We use the following normalization for the
polarization tensor:
\be
\bar \Psi^{(i)}_{\mu_1 \ldots\mu_n}
\Psi^{(j)}_{\mu_1 \ldots\mu_n}=(-1)^n \delta_{ij}\,.
\label{norm_psi}
\ee
With this choice of the sign, the tensor part of the propagator
differs from that introduced in \cite{operators}
by a factor $(-1)^n$. However, the present definition is more
convenient to calculate the unitarity condition, and we keep it in
further calculations.

Multiplying the amplitude squared by $\Psi^{(i)}_{\alpha_1
\ldots\alpha_n} \bar \Psi^{(i)}_{\beta_1 \ldots\beta_n}$ and summing
over polarizations we obtain:
\be
&&F^{\alpha_1\ldots\alpha_n}_{\beta_1\ldots\beta_n}\!
M\!\Gamma\!=\!\!\int\!\!\frac{(2\pi)^4}{2}
d\Phi_m(P,q_1\ldots q_m)\,g^2(s)\times
\nn
&&\sum\limits_{i=1}^{2J+1}\!\!
\Psi^{(i)}_{\alpha_1 \ldots\alpha_n}
\bar \Psi^{(i)}_{\mu_1 \ldots\mu_n}
Q_{\mu_1\ldots\mu_n}\!\!\otimes\!
Q_{\nu_1\ldots\nu_n}
\Psi^{(i)}_{\nu_1 \ldots\nu_n}
\bar \Psi^{(i)}_{\beta_1 \ldots\beta_n}.~~ \label{add1}
\ee
Here, the expression
$Q_{\mu_1\ldots\mu_n}\otimes Q_{\nu_1\ldots\nu_n}$ assumes
summation over polarizations of the final particles. Due to
orthogonality of the polarization tensors,
\be
\int\!\!\bar \Psi^{(i)}_{\mu_1 \ldots\mu_n}\!
Q_{\mu_1\ldots\mu_n}\!\!\!\otimes\!
Q_{\nu_1\ldots\nu_n}\!
\Psi^{(j)}_{\nu_1 \ldots\nu_n}
d\Phi_m(P,q_1\ldots q_m)
\!\sim\!\delta_{ij}~
\ee
the product of the polarization tensors can be
substituted by
\be
\Psi^{(i)}_{\nu_1 \ldots\nu_n} \bar \Psi^{(i)}_{\beta_1 \ldots\beta_n}
\to
\sum\limits_{j=1}^{2J+1}\!
\Psi^{(j)}_{\nu_1 \ldots\nu_n} \bar \Psi^{(j)}_{\beta_1 \ldots\beta_n}=
F^{\nu_1 \ldots\nu_n}_{\beta_1 \ldots\beta_n}\,.
\ee
Finally, we obtain
\be
F^{\alpha_1\ldots\alpha_n}_{\beta_1\ldots\beta_n}\,
M\!\Gamma\!&=&\!\int\!\frac{(2\pi)^4}{2}
d\Phi_m(P,q_1\ldots q_m)\,g^2(s) \times
\nn
&&F^{\alpha_1 \ldots\alpha_n}_{\mu_1 \ldots\mu_n}
Q_{\mu_1\ldots\mu_n}\otimes
Q_{\nu_1\ldots\nu_n}
F^{\nu_1 \ldots\nu_n}_{\beta_1 \ldots\beta_n}\,.~~~~
\label{width}
\ee
This is the basic equation  for the calculation of partial widths of
resonances. Another form of this equation can be obtained after the
convolution of both sides of equation (\ref{width}) with the
metric tensors $g_{\alpha_1\beta_1}\ldots g_{\alpha_n\beta_n}$ and
taking into account the properties (\ref{psi_psi}) and (\ref{norm_psi}):
\be
(2J\!+\!1)M\!\Gamma&=&\int\!\frac{(2\pi)^4}{2}
d\Phi_m(P,q_1\ldots q_m)g^2(s) \times
\nn
&&Q_{\mu_1\ldots\mu_n}\!\otimes\!
Q_{\nu_1\ldots\nu_n}
F^{\nu_1 \ldots\nu_n}_{\mu_1 \ldots\mu_n}\;.
\label{width_2}
\ee
Then the cross section for $2\to m$ particle transition can be rewritten as
\be
&&\sigma=\int
\frac{(2\pi)^4}{4|\vec k|\sqrt{s}}\,
d\Phi_m(P,q_1\ldots q_m)\,g_{in}^2
Q^{in}_{\mu_1\ldots\mu_n} \times
\nn
&&\frac{F^{\mu_1 \ldots\mu_n}_{\nu_1\ldots\nu_n}
Q^{out}_{\nu_1\ldots\nu_n}\otimes Q^{out}_{\alpha_1\ldots\alpha_n}
F^{\alpha_1 \ldots\alpha_n}_{\beta_1\ldots\beta_n}}
{(M^2-s)^2+(M\Gamma_{tot})^2}\,
Q^{in}_{\beta_1\ldots\beta_n}\,g_{out}^2=
\nn
&&\frac{g^2_{in}}{2|\vec k|\sqrt{s}}
\frac{Q^{in}_{\mu_1\ldots\mu_n}
F^{\mu_1 \ldots\mu_n}_{\beta_1\ldots\beta_n}M\Gamma_{out}
Q^{in}_{\beta_1\ldots\beta_n}}
{(M^2-s)^2+(M\Gamma_{tot})^2}\,.~~~~
\ee
The two--body phase space of particles with masses $m_1$ and $m_2$
is equal to
\be
\frac{(2\pi)^4}{2} d\Phi_2(P,k_1,k_2)=
\rho(s,m_1,m_2)\frac{d\Omega}{4\pi}\,,
\label{phv_2b0}
\ee
where the invariant form of the $\rho$ function is given by
\be
\rho(s,m_1,m_2)\!=\!\frac{\sqrt{(s\!-\!(m_1\!+\!m_2)^2))
(s\!-\!(m_1\!-\!m_2)^2)}}{16\pi s}\,.~~
\label{phv_2b}
\ee
This function can be also expressed through momentum of the
particles in the c.m.s of the reaction:
\be
\rho(s,m_1,m_2)=\frac{1}{16\pi}\frac{2|\vec k|}{\sqrt s}\,.
\ee
Then, we can use eq.(\ref{width_2}) to calculate partial width for
decays into the initial--state particles. If they have spins $s_1$ and
$s_2$, the cross section is calculated by averaging over their
polarizations.
After the summation over spin variables, one has for the partial width:
\be
\sigma=\frac{2J+1}{(2s_1\!+\!1)(2s_2\!+\!1)}\frac{4\pi}{|\vec k|^2}
\frac{M^2\Gamma_{in} \Gamma_{out}}
{(M^2\!-\!s)^2\!+\!(M\Gamma_{tot})^2}\;.
\ee
It is the standard equation for the contribution of a resonance with
spin $J$ to the cross section.

\section{The width of Breit-Wigner resonance}

\subsection{Rescatering of particles and Breit-Wigner states}

\begin{figure}[ht]
\centerline{\epsfig{file=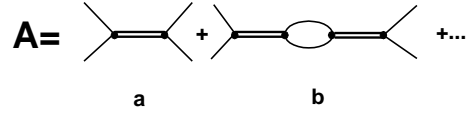,width=8.cm,clip=on}}
\caption{Resonance rescattering diagrams}
\label{res_prod}
\end{figure}

The amplitude for the rescattering of two spinless particles with total
momentum $P$ ($s=P^2$) via scalar resonance with
the bare mass $M_0$ (Fig.~\ref{res_prod}) can be written as
an infinite sum:
 \be
 A(s)\! =\!\frac{g^2(s)}{M_0^2-s}\!+\!
 \frac{g(s)}{M_0^2-s}B(s)\frac{g(s)}{M_0^2-s}\!+\!\ldots
\label{loops_sim}
 \ee
where the $B(s)$ function corresponds to the loop in the intermediate state. It can be
calculated, for example, in terms of the Feynman integral. However, for
our purpose it is convenient to use the dispersion relation technique where the
real part of the amplitude can be constructed as a dispersion integral
over the imaginary part. The imaginary part of the loop diagram is
equal to

\be
Im B(s) = g^2(s) \int\frac{(2\pi)^4}{2} d\Phi_2(P,k_1,k_2)
\ee
where $k_i$ and $m_i$ are the momenta and masses of the particles in the
loop. Then, $B(s)$ can be written as the integral
\be
B(s)= \int\limits_{(m_1+m_2)^2}^\infty \frac{ds'}{\pi}
\frac{Im\,B(s')}{s'-s -i0}\;.
\ee
In the considered spinless particle
case, the amplitude corresponding to
(\ref{loops_sim}) can be   easily calculated
\be
A(s)\! =\!\frac{g^2(s)}{M_0^2-s }\;\frac{1}{1-\frac{B(s)}{M_0^2-s}}
\! =\!\frac{g^2(s)}{M_0^2-s-B(s)}\,.
\ee
The real part of the $B(s)$ function has left-hand singularities in the
complex-$s$ plane and is a smooth function in the physical region.
Therefore in the resonance region
\be
Re B(s) \simeq Re B(M^2) + Re B'(M^2)(s-M^2)\,,
\ee
and the amplitude can be written as
\be
A(s)\! =\!\frac{\tilde g^2(s)}{M^2-s- iM \Gamma}\,,
\ee
where
\be
&&M^2\!=M_0^2-Re B(M^2) \;, \nn
&&M\Gamma\!=Im B(M^2)/(1+Re B'(M^2)) \;,\nn
&&\tilde g^2(s) = g^2(s)/(1+Re B'(M^2))
\label{new1}
\ee
The first line of eq.(\ref{new1}) defines the position of the resonance,
the imaginary part of $B(s)$ defines the width of the state.

Now consider the general case. The amplitude which describes
the scattering via a resonance with total spin $J$ is given by
\be
&&A(s)\!=\!
g_{in}\,Q^{in}_{\mu_1\ldots\mu_n} \,
\frac{F^{\mu_1\ldots\mu_n}_{\nu_1\ldots\nu_n}}{M_0^2-s}\,
Q^{out}_{\nu_1\ldots\nu_n}\,g_{out}+
\\
&&g_{in}Q^{in}_{\mu_1\ldots\mu_n} \,
\frac{F^{\mu_1\ldots\mu_n}_{\nu_1\ldots\nu_n}}{M_0^2-s}\,
\tilde B^{\nu_1\ldots\nu_n}_{\xi_1\ldots\xi_n}
\frac{F^{\xi_1\ldots\xi_n}_{\beta_1\ldots\beta_n}}{M_0^2-s}\,
Q^{out}_{\beta_1\ldots\beta_n}\,g_{out}
\!+\!\ldots~~
\nonumber
\label{bw_amp}
\ee
Here, $F^{\mu_1\ldots\mu_n}_{\nu_1\ldots\nu_n}$ is the tensor part of
the propagator, $Q^{in}$, $Q^{out}$ are vertex operators for the initial
and final states ($n=J$ for boson and $n=J\!-\!1/2$ for
fermion resonances). We assume also that the vertex operators include
the polarization tensors of the initial and final particles.

The imaginary part of the loop diagram for the intermediate state
with $m$ particles is given by
\be
Im \tilde B^{\nu_1\ldots\nu_n}_{\xi_1\ldots\xi_n}\!&=&\!
\int\!\!\frac{(2\pi)^4}{2} g^2(s)
d\Phi_{m}
Q_{\nu_1\ldots\nu_n}\!\!\otimes\!Q_{\xi_1\ldots\xi_n}\,,
\nn
~~~~d\Phi_{m}&\equiv& d\Phi_{m}(P,k_1,\ldots,k_{m})\,.
\ee
Here, $g(s)$ and $Q$ are the coupling and vertex operator,
respectively, for the decay of a resonance into the intermediate state.
As before, the definition
$Q_{\nu_1\ldots\nu_n}\!\otimes\!Q_{\xi_1\ldots\xi_n}$ assumes summation
over polarizations of the intermediate particles. In the pure elastic
case, the intermediate state operator is equal to $Q=Q^{in}=Q^{out}$
but generally, the $B$--function is equal to the sum of loop
diagrams over all possible decay modes.

Let us define the $B(s)$-function  as follows:
\be
F^{\mu_1\ldots\mu_n}_{\beta_1\ldots\beta_n}Im B(s)=
F^{\mu_1\ldots\mu_n}_{\nu_1\ldots\nu_n}
Im \tilde B^{\nu_1\ldots\nu_n}_{\xi_1\ldots\xi_n}
F^{\xi_1\ldots\xi_n}_{\beta_1\ldots\beta_n}\!=
\nn
F^{\mu_1\ldots\mu_n}_{\nu_1\ldots\nu_n}\!\!
\int\!\!\frac{(2\pi)^4}{2} g^2(s)
d\Phi_{m}\,
Q_{\nu_1\ldots\nu_n}\!\otimes\!Q_{\xi_1\ldots\xi_n}
F^{\xi_1\ldots\xi_n}_{\beta_1\ldots\beta_n}
\label{width_bw}
\ee
Using this equation, one can convolute all tensor factors into
one structure, so one obtains:
\be
A(s)\!=\!
g_{in}\,Q^{in}_{\mu_1\ldots\mu_n} \,
\frac{F^{\mu_1\ldots\mu_n}_{\nu_1\ldots\nu_n}}{M_0^2-s}\,
Q^{out}_{\nu_1\ldots\nu_n}\,g_{out}\Big( 1+
\nn
\frac{B(s)}{M_0^2-s}+\frac{B(s)}{M_0^2-s}\frac{B(s)}{M_0^2-s}+
\ldots\Big )=
\nn
g_{in}\,Q^{in}_{\mu_1\ldots\mu_n} \,
\frac{F^{\mu_1\ldots\mu_n}_{\nu_1\ldots\nu_n}}{M_0^2-s-B(s)}\,
Q^{out}_{\nu_1\ldots\nu_n}\,g_{out}\,.
\label{bw_fin}
\ee
As before, the imaginary part of the B-function defines the width of
the state, and we obtain the standard Breit-Wigner expression.

\subsection{Decay of the resonance into two spinless particles}

For the rescattering of two spinless particles the vertex is
defined by the orbital momentum operators only. These operators satisfy
the symmetry, orthogonality and traceless properties (see
\cite{oper_orig} for more detail):
\be
&&X^{(L)}_{\mu_1\ldots\mu_i\ldots\mu_j\ldots\mu_L}\; =\;
X^{(L)}_{\mu_1\ldots\mu_j\ldots\mu_i\ldots\mu_L}.
\nn
&&P_{\mu_i}X^{(L)}_{\mu_1\ldots\mu_i\ldots\mu_L}\ =\ 0.
\nn
&&g_{\mu_i\mu_j}X^{(L)} _{\mu_1\ldots\mu_i\ldots\mu_j\ldots\mu_L}\
\ =\ 0,
\label{x_oper}
\ee
where $L$ is the orbital momentum. The second property means that the
operators $X$ are constructed from vectors and tensors orthogonal to
the total momentum. In the case of the two--particle final state, only
one vector of such a type can be constructed:
\be k^\perp_\mu=\frac12
g^\perp_{\mu\nu}(k_1-k_2)_\nu \;, \qquad
g^\perp_{\mu\nu}=g_{\mu\nu}-\frac{P_\mu P_\nu}{s}\;,~~~
\ee
where $k_1$ and $k_2$ are the momenta of the constituents.

The orbital angular momentum operators for $L \le 3 $ are:
\begin{eqnarray}
X^{(0)}&=&1\ , \qquad X^{(1)}_\mu=k^\perp_\mu\ , \qquad\nonumber \\
X^{(2)}_{\mu_1 \mu_2}&=&\frac32\left(k^\perp_{\mu_1}
k^\perp_{\mu_2}-\frac13\, k^2_\perp g^\perp_{\mu_1\mu_2}\right), \nonumber  \\
X^{(3)}_{\mu_1\mu_2\mu_3}&=&\frac52\Big[k^\perp_{\mu_1} k^\perp_{\mu_2 }
k^\perp_{\mu_3}
\nn
&-&\frac{k^2_\perp}5\left(g^\perp_{\mu_1\mu_2}k^\perp
_{\mu_3}+g^\perp_{\mu_1\mu_3}k^\perp_{\mu_2}+
g^\perp_{\mu_2\mu_3}k^\perp_{\mu_1}
\right)\Big]\,.~~~
\ee
The operators $X^{(L)}_{\mu_1\ldots\mu_L}$ for $L\ge 1$ can be written
in the form of the recurrency expression:
\be
X^{(L)}_{\mu_1\ldots\mu_L}&=&k^\perp_\alpha
Z^{\alpha}_{\mu_1\ldots\mu_L} \; ,
\nonumber\\
Z^{\alpha}_{\mu_1\ldots\mu_L}&=&
\frac{2L-1}{L^2}\Big (
\sum^L_{i=1}X^{{(L-1)}}_{\mu_1\ldots\mu_{i-1}\mu_{i+1}\ldots\mu_L}
g^\perp_{\mu_i\alpha}-
\nonumber \\
 \frac{2}{2L-1}  \sum^L_{i,j=1 \atop i<j}
&g^\perp_{\mu_i\mu_j}&
X^{{(L-1)}}_{\mu_1\ldots\mu_{i-1}\mu_{i+1}\ldots\mu_{j-1}\mu_{j+1}
\ldots\mu_L\alpha} \Big )\,.
%\nonumber
\label{z}
\ee
Other useful properties of the orbital momentum operators are listed in
Appendix.

The projection operator $O^{\mu_1\ldots\mu_L}_{\nu_1\ldots \nu_L}$
is constructed from the metric tensors $g^\perp_{\mu\nu}$ and
has the following properties:
\be
X^{(L)}_{\mu_1\ldots\mu_L}
O^{\mu_1\ldots\mu_L}_{\nu_1\ldots \nu_L}\
&=&\ X^{(L)}_{\nu_1\ldots \nu_L}\ , \nonumber \\
O^{\mu_1\ldots\mu_L}_{\alpha_1\ldots\alpha_L} \
O^{\alpha_1\ldots\alpha_L}_{\nu_1\ldots \nu_L}\
&=& O^{\mu_1\ldots\mu_L}_{\nu_1\ldots \nu_L}\ .
\label{proj_op}
\ee
The projection operator projects any tensor with $n$ indices onto
tensors which satisfy the properties (\ref{x_oper}).
For the lowest states,
\be
O\!&=&\! 1\qquad O^\mu_\nu\!=\!g_{\mu\nu}^\perp
\nn
O^{\mu_1\mu_2}_{\nu_1\nu_2}\!&=&\!
\frac 12 \left (
g_{\mu_1\nu_1}^\perp  g_{\mu_2\nu_2}^\perp \!+\!
g_{\mu_1\nu_2}^\perp  g_{\mu_2\nu_1}^\perp  \!- \!\frac 23
g_{\mu_1\mu_2}^\perp  g_{\nu_1\nu_2}^\perp \right )\,.~
\ee
For higher states, the operator can be calculated using the
recurrent expression:
\be &&O^{\mu_1\ldots\mu_L}_{\nu_1\ldots
\nu_L}=\frac{1}{L^2} \bigg (
\sum\limits_{i,j=1}^{L}g^\perp_{\mu_i\nu_j}
O^{\mu_1\ldots\mu_{i-1}\mu_{i+1}\ldots\mu_L}_{\nu_1\ldots
\nu_{j-1}\nu_{j+1}\ldots\nu_L}-
\nonumber \\
 &&  \frac{4}{(2L-1)(2L-3)} \times
\nn    &&
\sum\limits_{i<j\atop k<m}^{L}
g^\perp_{\mu_i\mu_j}g^\perp_{\nu_k\nu_m}
O^{\mu_1\ldots\mu_{i-1}\mu_{i+1}\ldots\mu_{j-1}\mu_{j+1}\ldots\mu_L}_
{\nu_1\ldots\nu_{k-1}\nu_{k+1}\ldots\nu_{m-1}\nu_{m+1}\ldots\nu_L}
\bigg )\,.
\ee
The tensor part of the boson propagator is defined by the projection
operator. Let us write it as
\be
F^{\mu_1\ldots\mu_L}_{\nu_1\ldots\nu_L}=
(-1)^L\,O^{\mu_1\ldots\mu_L}_{\nu_1\ldots \nu_L}\,.
\label{boson_prop}
\ee
This definition differs from the expression given in \cite{operators}
by the factor $(-1)^L$. This choice simplifies the
expressions for amplitudes given in \cite{operators} where this factor
was explicitly included  in all final expressions. Furthermore,
this definition guarantees that the width of the resonance (when it is
calculated from the vertices) is a positive value.

The production of the two $X$-operators integrated over solid
angle (which is equivalent to the integration over internal momenta)
depends on external momenta and the metric tensor only. Therefore, it
must be proportional to the projection operator.
After straightforward calculations, we obtain:
\be
\int\!\frac{d\Omega }{4\pi}\!
X^{(L)}_{\mu_1\ldots\mu_L}(k^\perp)
X^{(L)}_{\nu_1\ldots\nu_L}(k^\perp)\!=\!
 \frac{\alpha_L\,k^{2L}_\perp}{2L\!+\!1}
O^{\mu_1\ldots\mu_L}_{\nu_1\ldots \nu_L},~~
\label{x-prod}
\ee
where
\be
\alpha_L=\prod^L_{l=1}\frac{2l-1}{l}\; .
\label{alpha}
\ee
The width of the state is calculated by means of
eq.(\ref{width}) using the properties (\ref{proj_op}):
\be
M\Gamma
\!=\!(-1)^L\;\frac{\alpha_L g^2(s)}{2L+1}k_\perp^{2L}\,\rho(s,m_,m_2)\,,~
\label{bw_1}
\ee
where $\rho(s,m_1,m_2)$ is defined by eqs.(\ref{phv_2b0},\ref{phv_2b})
and $g(s)$ is the coupling.

Let us introduce the positive value $|\vec k|^2$:
\be
|\vec k|^2\!=\!-k_\perp^2\!=\!
\frac{[s\!-\!(m_1\!+\!m_2)^2][s\!-\!(m_1\!-\!m_2)^2]}{4s}\,.
\label{k2_rel}
\ee
In the c.m.s. of the reaction, $\vec k$ corresponds to the momentum of
the particle. In other systems we use this definition only in the sense
of $|\vec k|\equiv \sqrt{-k_\perp^2}$, and therefore it is
relativistic invariant positive value. Then we obtain
\be M\Gamma \!=\!\frac{\alpha_L
g^2(s)}{2L+1}|\vec k|^{2L}\,\rho(s,m_1,m_2)~~
\label{boson_width} \ee
as the width of a state. The amplitude for the rescattering of
constituents with the initial relative momentum $k$ and final relative
momentum $q$ is equal to
\be
A(s)
\! &=&\! g(s)
X^{(L)}_{\mu_1\ldots\mu_L}(k^\perp)
\frac{(-1)^L\;O^{\mu_1\ldots\mu_l}_{\nu_1\ldots\nu_L}}{M^2-s-i M\Gamma}
X^{(L)}_{\nu_1\ldots\nu_L}(q^\perp) g(s)
\nn
\!& =&\! g(s)
\frac{\alpha_L P_L(z)(|\vec k||\vec q|)^L}
{M^2-s-i M\Gamma}\,g(s)\;.~~~~~~
\label{amp_bos}
\ee
Here $P_L(z)$ are Legendre polynomials and $z$ is equal to
\be
z=\frac{k^\perp q^\perp}{\sqrt{k^2_\perp}\sqrt{q^2_\perp}}\;.
\ee
In the c.m.s. of the resonance, $z$ is the cosine of the angle
between the momenta of initial--state and final--state particles.
To obtain eq.(\ref{amp_bos}), we used the following property of the
$X$-operators:
\be
X^{(L)}_{\mu_1\ldots\mu_L}(k^\perp)X^{(L)}_{\mu_1\ldots\mu_L}
(q^\perp)\!=\! \alpha_L\!\!
\left(\sqrt{k^2_\perp}\sqrt{ q^2_\perp}\right)^{L}\!\! P_L(z)\,.~~
\label{kq_ampl}
\ee
The total cross section for the elastic scattering is equal to:
\be
\sigma\!&=&\! \frac{(2\pi)^4}{4|\vec k|\sqrt{s}} \!\int\!
\frac{g^4(s)\alpha^2_L P^2_L(z)(|\vec k||\vec q|)^{2L}}
{(M^2-s)^2+(M\Gamma)^2}
d\Phi_2(P,q_1,q_2)
\nn
&=&\!\frac{4\pi}{|\vec k|^2}(2L\!+\!1)
\frac{(M\Gamma)^2}
{(M^2\!-\!s)^2\!+\!(M\Gamma)^2}\,.
\ee

%=============================================================================================

\section{Partial widths of the baryon resonances}

\subsection{The structure of the fermion propagator}

The structure of the fermion propagator
$\mathcal P^{\mu_1\ldots\mu_n}_{\nu_1\ldots\nu_n}$
 was considered in details in
\cite{operators}. The propagator is defined as
\be
\mathcal P^{\mu_1\ldots\mu_n}_{\nu_1\ldots\nu_n}\!=
\frac{F^{\mu_1\ldots\mu_n}_{\nu_1\ldots\nu_n}}{M^2 -s -iM\Gamma}\,,
\ee
where
\be
F^{\mu_1\ldots\mu_n}_{\nu_1\ldots\nu_n}\!=\!(-1)^n
\frac{\sqrt{s}\!+\!\hat P}{2\sqrt{s}}
O^{\mu_1\ldots\mu_n}_{\xi_1\ldots \xi_n}
T^{\xi_1\ldots\xi_n}_{\beta_1\ldots \beta_n}
O^{\beta_1\ldots \beta_n}_{\nu_1\ldots\nu_n}\,.
\label{fp}
\ee
Here, $(\sqrt s+\hat P)$ corresponds to the numerator of fermion
propagator describing the particle with $J=1/2$ and
$n\!=\!J\!-\!1/2$ ($\sqrt s\!=\!M$ for the stable particle). We
define
\be T^{\xi_1\ldots\xi_n}_{\beta_1\ldots \beta_n}&=&
\frac{n+1}{2n\!+\!1} \big( g_{\xi_1\beta_1}\!-\!
\frac{n}{n\!+\!1}\sigma_{\xi_1\beta_1} \big)
\prod\limits_{i=2}^{n}g_{\xi_i\beta_i}, \nn
\sigma_{\alpha_i\alpha_j}&=&\frac 12
(\gamma_{\alpha_i}\gamma_{\alpha_j}-
\gamma_{\alpha_j}\gamma_{\alpha_i}).
\label{t1}
\ee
As in \cite{operators}, we introduced the factor $1/(2\sqrt s)$ in the
propagator which removes the divergency of this function at large
energies. For the stable particle it means that bispinors are normalized
as follows:
\be \bar u(k_N) u(k_N)\!=\!1\;,\;\;
\sum\limits_{polarizations}\!\!\!\!\!\! u(k_N)\bar u(k_N)
\!=\!\frac{m\!+\!\hat k_N}{2m}\;.
\label{bisp_norm}
\ee
Here and below, $\hat k\equiv\gamma_\mu k_\mu$.

It is useful to list the properties of the fermion propagator:
\be
&&P_{\mu_i}F^{\mu_1\ldots\mu_n}_{\nu_1\ldots \nu_n}
=P_{\nu_j}F^{\mu_1\ldots\mu_n}_{\nu_1\ldots \nu_n}=0\;,
\nn
&&\gamma_{\mu_i}F^{\mu_1\ldots\mu_n}_{\nu_1\ldots \nu_n}
=F^{\mu_1\ldots\mu_n}_{\nu_1\ldots \nu_n}\gamma_{\nu_j}=0\;,
\nn
&&F^{\mu_1\ldots\mu_n}_{\alpha_1\ldots \alpha_n}
F^{\alpha_1\ldots \alpha_n}_{\nu_1\ldots \nu_n}=
(-1)^n F^{\mu_1\ldots\mu_n}_{\nu_1\ldots \nu_n} \;,
\nn
&&\hat P F^{\mu_1\ldots\mu_n}_{\nu_1\ldots \nu_n}
=\sqrt s F^{\mu_1\ldots\mu_n}_{\nu_1\ldots \nu_n}.
\label{F_proper}
\ee
%=============================================================================================

\subsection{\boldmath $\pi N$ partial widths of baryon resonances}

The operators which describe the decay of a baryon into
the $\pi N$
system were introduced in \cite{operators}.
The states with $J=L\!+\!1/2$, where $L$ is the orbital momentum of the
$\pi N$ system, are called '+' states
($1/2^-$, $3/2^+$, $5/2^-$,\ldots).
The states with $J=L\!-\!1/2$ are called '-' states
($1/2^+$, $3/2^-$, $5/2^+$,\ldots).
The correspondent vertices are ($n=J\!-\!1/2$):
\be
N^+_{\mu_1\ldots\mu_n}(k^\perp)u(k_N)\!&=&\!
X^{(n)}_{\mu_1\ldots\mu_n}(k^\perp)u(k_N)\,.
\nn
N^-_{\mu_1\ldots\mu_{n}}(k^\perp)u(k_N)\!&=&\!
i\gamma_5 \gamma_\nu
X^{(n+1)}_{\nu\mu_1\ldots\mu_{n}}(k^\perp)u(k_N) \,.
\ee
Here, $u(k_N)$ is the bispinor of the final--state nucleon.

The width is defined by the equation (\ref{width}) which
for the case of $\pi N$ scattering has the form
\be
&&F^{\mu_1\ldots\mu_n}_{\nu_1\ldots \nu_n}M\Gamma^{\pm}_{\pi N}=
F^{\mu_1\ldots\mu_n}_{\xi_1\ldots \xi_n}
\int\frac{d\Omega}{4\pi}
\tilde N^\pm_{\xi_1\ldots\xi_n}\times
\nn
&&\frac{\hat k_{N}\!+\!m_N}{2m_N}
N^\pm_{\beta_1\ldots\beta_n}\rho(s,m_\pi,m_N)g^2(s)
 F^{\beta_1\ldots\beta_n}_{\nu_1\ldots \nu_n}.~~~~
\label{width_piN}
\ee
Here, we use $\tilde N$ to define the vertex which is different
from $N$ by the order of gamma matrices.

 The momentum of the nucleon can be decomposed in the total
 momentum P and momentum $k^\perp$ as:
 \be
 k_{N\mu} = \frac{k_{N0}}{\sqrt{s}}P_\mu + k_\mu^\perp,\;\;
 k_{N0}=\frac{s+m^2_N - m^2_{\pi}}{2\sqrt s}
 \label{k0n}
 \ee
where
\be
k_\mu^\perp = \frac12(k_N - k_\pi)_\mu -
\frac{m_N^2 -m_\pi^2}{2s} P_\mu\,.
\ee

For '+' states the calculation can be easily performed using
eq.(\ref{x-prod}), the last property from eqs. (\ref{F_proper}) and
the condition that integral over the odd number of $k^\perp_\mu$ vectors
vanishes. Then,
\be
M \Gamma^{+}_{\pi N}\!=\! \frac {\alpha_n}{2n+1}|\vec k|^{2n}
 \frac{m_N\!+\!k_{N0}}{2m_N} \rho(s,m_\pi,m_N)g^2(s)\,.~~
\label{res_piN_plus}
\ee
For the '-' states the calculations are more complicated.
Using the formulae given in Appendix, one obtains finally
\be
M\Gamma^-_{\pi N} =
\frac {\alpha_{n+1}}{n+1}|\vec k|^{2n+2} \frac{m_N\!+\!k_{N0}}
{2m_N} \rho(s,m_\pi,m_N)g^2(s)\,.~~
\label{res_piN_minus}
\ee

The definition of the propagator in the form (\ref{fp}) provides a
positive magnitude of the resonance widths and correct
position of the poles of the scattering amplitude. Let us write
with this definition the expression for the scattering amplitude.
The partial amplitude for the $\pi N$ scattering from the initial state
with relative momentum $k^\perp$ into the final state with momentum
$q^\perp$ is defined by:
\be
A_{\pi N}\!&=&\!\sum\limits_{n} A_n^{+} BW^{+}_n(s)+ A_n^{-}
BW^{-}_n(s)\,,
\nn
A_n^{\pm}\! &=&\bar u(k_1)\tilde
N^{\pm}_{\mu_1\ldots\mu_n}(k^\perp\!)
F^{\mu_1\ldots\mu_n}_{\nu_1\ldots\nu_n}
N^{\pm}_{\nu_1\ldots\nu_n}(q^\perp)
u(q_1)\,.~~~~
\ee
Here, $BW^{\pm}_n(s)$ is the energy-dependent part of the amplitude.

In the c.m.s. of the reaction (see \cite{operators} for more detail),
this amplitude can be rewritten as
\be
&&A_{\pi N}=\omega^*\left [G(s,t)+H(s,t)i(\vec \sigma \vec n)
\right ]\omega' \;,
\nonumber \\
&&G(s,t)=\sum\limits_L \big [(L\!+\!1)F_L^+(s)- L F_L^-(s)\big ]
P_L(z) \;, \nonumber \\
&&H(s,t)=\sum\limits_L \big
[F_L^+(s)+ F_L^-(s)\big ] P'_L(z) \;,
\label{piN_others}
\ee
where $\omega$ and $\omega'$
are nonrelativistic spinors and $\vec n$ is a unit vector
normal to the decay plane. The $F$-functions are defined as follows:
\be
F^+_L&=&(|\vec k||\vec q|)^L
\sqrt{\chi_i\chi_f}\;\frac{\alpha_L}{2L\!+\!1} BW_L^+(s) \;,
\nonumber \\
F^-_L&=&(|\vec k||\vec q|)^L
\sqrt{\chi_i\chi_f}\;\frac{\alpha_L}{L} BW_L^-(s)\,,
\nn
\chi_i&=&m_N+k_{N0}\;, \qquad \chi_f=m_N+q_{N0}\;,
\ee
where $L=n$ stands for '+' states and $L=n+1$ for '-' states.

%=============================================================================================

\section{The $\gamma N$ widths and helicity amplitudes}

The decay of the state with $J=n+1/2$ into $\gamma N$ is described by the
amplitude:
\be
\bar \Psi_{\alpha_1\ldots\alpha_n}
V^{(i\pm)\mu}_{\alpha_1\ldots\alpha_n}(k^\perp)
 u(k_N)  \varepsilon_\mu \;,  \nonumber
\ee
where $k_N$ is the momentum of the nucleon and
$k^\perp$ is the component of the relative momentum between nucleon and
photon which is orthogonal to the total momentum of the system $P$
($s=P^2$):
\be k^\perp_\mu=\frac12 &&(k_N-k_\gamma)_\nu
g^\perp_{\mu\nu}  \;, \qquad
g^\perp_{\mu\nu}=g_{\mu\nu}-\frac{P_\mu P_\nu}{s}\;,~~~
\nn
&&|\vec k|^2=-k_\perp^2=\frac{(s-m_N)^2}{4s}\,.
\ee

\subsection{The '+' states}

For the states with $n\ge 1$, three vertices can be constructed of the
spin and orbital momentum operators. For '+' states the vertices are:
\be
&&V^{(1+)\mu}_{\alpha_1\ldots\alpha_n}(k^\perp)=
\gamma^\perp_\mu i\gamma_5
X^{(n)}_{\alpha_1\ldots\alpha_n}(k^\perp) \;,\nonumber\\
&&V^{(2+)\mu}_{\alpha_1\ldots\alpha_n}(k^\perp)=
\gamma_\nu i \gamma_5
X^{(n+2)}_{\mu\nu\alpha_1\ldots\alpha_n}(k^\perp) \nonumber \;,\nn
&&V^{(3+)\mu}_{\alpha_1\ldots\alpha_n}(k^\perp)=
\gamma_\nu i \gamma_5
X^{(n)}_{\nu\alpha_1\ldots\alpha_{n-1}}(k^\perp)
g^\perp_{\mu\alpha_n} \;.
\label{vf_plus}
\ee
The first vertex is constructed using the spin $1/2$ operator and
$L=n$ orbital momentum operator, the second one has $S=3/2$, $L=n+2$
and the third one $S=3/2$ and $L=n$. In case of
photoproduction, the second vertex is reduced to the third one and only
two amplitudes (one for $J=1/2$) are independent. The width factor
$W^{(i,j+)}$for the transition between vertices is expressed as
follows:
\be
F^{\mu_1\ldots\mu_n}_{\nu_1\ldots \nu_n} W^+_{i,j}=
F^{\mu_1\ldots\mu_n}_{\alpha_1\ldots \alpha_n} \int
\!\frac{d\Omega}{4\pi} \tilde
V^{(i+)\mu}_{\alpha_1\ldots\alpha_n}(k^\perp) \times
 \nn \frac{m_N\!+\!\hat
k_N}{2m_N} V^{(j+)\nu}_{\beta_1\ldots\beta_n}(k^\perp)
(-g_{\mu\nu}^{\perp\perp})\rho(s)
F^{\beta_1\ldots\beta_n}_{\nu_1\ldots\nu_n}\;,
\label{gN_general}
\ee
where $\rho(s)\!\equiv\!\rho(s,m_N,m_\gamma)$,
$(-g_{\mu\nu}^{\perp\perp})$ describes the structure of the
photon propagator:
\be
-g_{\mu\nu}^{\perp\perp}=
-g_{\mu\nu}+\frac{P_\mu P_\nu}{P^2}+
\frac{k^\perp_\mu k^\perp_\nu}{k_\perp^2}\;.
\label{g_2p}
\ee
and the operator $\tilde V$ differs from the operator $V$ by
the ordering of $\gamma$-matrices.
The width factors $W^{(i,j+)}$ for the first and
third vertices are equal to:
\be
W^+_{1,1}&=&\frac{2\alpha_n}{2n+1}|\vec k|^{2n}
\frac{m_N\!+\!k_{N0}}{2m_N}\rho(s)\;,
\nonumber\\
W^+_{3,3}&=&\frac{\alpha_n(n+1)}{(2n+1)n}
|\vec k|^{2n}\frac{m_N\!+\!k_{N0}}{2m_N}\rho(s)\;,
\nonumber \\
W^+_{1,3}&=&\frac{\alpha_n}{2n+1}|\vec
k|^{2n}\frac{m_N\!+\!k_{N0}}{2m_N}\rho(s)\,,
\label{w_plus}
\ee
where $\alpha_n$ is defined by (\ref{alpha}).

If a state with total spin $J\!=n+1/2$ decays into $\gamma N$ having
intrinsic spins $1/2$ and $3/2$ with couplings $g_1$ and $g_3$, the
corresponding decay amplitude can be written as
\be
A^{\mu(+)}_{\alpha_1\ldots\alpha_n}=
V^{(1+)\mu}_{\alpha_1\ldots\alpha_n}
g_1(s)+
V^{(3+)\mu}_{\alpha_1\ldots\alpha_n}
g_3(s)\,.
\ee
Then, the $\gamma N$ width is equal to:
\be
M\Gamma^+_{\gamma N}\!=
W^+_{1,1}\,g_1^2(s)\!+\!2W^+_{1,3}\,g_1(s)g_3(s)\!+\!
W^+_{3,3}\,g_3^2(s)\,.~~~
\ee
The helicity $1/2$ amplitude has an operator proportional to the spin
$1/2$ operator $V^{(1+)\mu}_{\alpha_1\ldots\alpha_n}$. The helicity
$3/2$ operator can be constructed as a linear combination of the spin
$3/2$ and $1/2$ operators, orthogonal to the
$V^{(1+)\mu}_{\alpha_1\ldots\alpha_n}$:
\be
A^{\mu(+)}_{\alpha_1\ldots\alpha_n}&=&
 A^{h=3/2}_{\mu;\alpha_1\ldots\alpha_n}
-A^{h=1/2}_{\mu;\alpha_1\ldots\alpha_n}\;,
\nonumber \\
A^{h=1/2}_{\mu;\alpha_1\ldots\alpha_n}&=&
-V^{(1+)\mu}_{\alpha_1\ldots\alpha_n}
\left ( g_1(s)+\frac 12 g_3(s)\right )\;,
\nonumber\\
A^{h=3/2}_{\mu;\alpha_1\ldots\alpha_n}&=&
\left (V^{(3+)\mu}_{\alpha_1\ldots\alpha_n}-\frac 12
V^{(1+)\mu}_{\alpha_1\ldots\alpha_n}\right )
g_3(s)\;,
\ee
where the sign "-" for the helicity $1/2$
amplitude was introduced in accordance with the
standard multipole definition.
The width defined by the helicity amplitudes can be calculated
using eq.(\ref{w_plus}).
\be
M\Gamma^{\frac 12}&=&\rho(s)W^+_{1,1}
\left (g_1(s)+\frac 12 g_3(s)\right )^2\;,
\nonumber \\
M\Gamma^{\frac 32}&=&\rho(s)
\left(W^+_{3,3}-\frac 12 W^+_{1,3} \right )
g_3^2(s)\;.
\ee
Taking into account the standard definition of the $\gamma N$ width
via helicity amplitudes,
\be
M\Gamma_{tot}\!=\!M\Gamma^{\frac 32}\!+\!M\Gamma^{\frac 12}\!=\!
\frac{\vec k^2}{\pi}\frac{2m_N}{2J\!+\!1}
\left ( |A_n^{\frac 12}|^2+|A_n^{\frac 32}|^2\right ),~~
\ee
we obtain
\be
|A_n^{\frac 12}|^2\!&=&\!\frac{\alpha_n(n\!+\!1)}{2n\!+\!1}\rho(s)
\pi|\vec k|^{2n-2}
\frac{\chi}{m^2_N}
\left (g_1(s)\!+\!\frac 12 g_3(s)\!\right )^2\,,
\nonumber \\
|A_n^{\frac 32}|^2\!&=&\!\alpha_n\rho(s)\pi|\vec k|^{2n-2}
\frac{\chi}{m^2_N}
\frac{(n\!+\!2)(n\!+\!1)}{4n(2n\!+\!1)}g_3^2(s)\,,
\label{hel_am_plus}
\ee
where $\chi=m_N+k_{N0}$.

In the case of resonance production, the vertex functions are
usually normalized with certain form factors, e.g. the Blatt-Weisskopf
form factors (the explicit form can be found in \cite{operators}).
These form factors depend on the orbital angular momentum and radius
$r$ and regularize the behaviour of the amplitude at large energies.
For the '+' states the orbital momentum for both spin 1/2 and 3/2
operators are equal to $L=J\!-\!1/2=n$. Then, rewriting
\be
g_1(s)=\frac{g_{1/2}}{F(n,|\vec
k|^2,r)},\qquad g_3(s)=\frac{g_{3/2}}{F(n,|\vec k|^2,r)},~~~~
\ee
and, using eq.(\ref{hel_am_plus}), the ratio of helicity
amplitudes given in \cite{operators} is reproduced.

\subsection{ The '-' states}

For the decay of a '-' state with total spin $J$ into $\gamma N$,
the vertex functions have the form:
\be
&&V^{(1-)\mu}_{\alpha_1\ldots\alpha_{n}}(k^\perp)=
\gamma_\xi\gamma^\perp_\mu
X^{(n+1)}_{\xi\alpha_1\ldots\alpha_{n}}(k^\perp) \;,
\nn
&&V^{(2-)\mu}_{\alpha_1\ldots\alpha_{n}}(k^\perp)=
X^{(n+1)}_{\mu\alpha_1\ldots\alpha_{n}}(k^\perp) \;,
\nonumber \\
&&V^{(3-)\mu}_{\alpha_1\ldots\alpha_{n}}(k^\perp)=
X^{(n-1)}_{\alpha_2\ldots\alpha_{n}}(k^\perp)
g^\perp_{\alpha_1\mu} \;.
\label{vf_minus}
\ee
These vertices are constructed of the spin and orbital momentum
operators with ($S=1/2$, $L=n+1$), ($S=3/2$, $L=n+1$) and
($S=3/2$ and $L=n-1$). As in case of "+" states, the second vertex
provides us the same angular distribution as the third vertex. For the
first and third vertices, the width factors $W^-_{i,j}$ are equal to
\be
W^-_{1,1}&=&\frac{2\alpha_{n+1}}{n+1}|\vec k|^{2n+2}
\frac{m_N\!+\!k_{N0}}{2m_N}\rho(s)\;,
\nonumber\\
W^-_{3,3}&=&\frac{\alpha_{n-1}(n+1)}{(2n\!+\!1)(2n\!-\!1)}
|\vec k|^{2n-2}\frac{m_N\!+\!k_{N0}}{2m_N}\rho(s)\;,
\nonumber \\
W^-_{1,3}&=&\frac{\alpha_{n-1}}{n+1}|\vec k|^{2n}
\frac{m_N\!+\!k_{N0}}{2m_N}\rho(s)\,,
\label{w_minus}
\ee
where $\rho(s)\!\equiv\!\rho(s,m_N,m_\gamma)$.

The decay amplitude is defined by the sum of two vertices as follows:
\be
A^{\mu(-)}_{\alpha_1\ldots\alpha_n}=
V^{(1-)\mu}_{\alpha_1\ldots\alpha_n}\,g_1(s)+
V^{(3-)\mu}_{\alpha_1\ldots\alpha_n}\,g_3(s)\;,
\ee
and the $\gamma N$ width of the state is calculated as a sum over
possible transitions:
\be
M\Gamma^-_{\gamma N}\!=
W^-_{1,1}\,g_1^2(s)\!+\!2W^-_{1,3}\,g_1(s)g_3(s)\!+\!
W^-_{3,3}\,g_3^2(s)\,.~~
\ee
The helicity $1/2$ amplitude has the operator proportional to the spin
$1/2$ operator $V^{(1+)\mu}_{\alpha_1\ldots\alpha_n}$. The helicity
$3/2$ operator can be constructed as a linear combination of the spin
$3/2$ and $1/2$ operators orthogonal to the
$V^{(1+)\mu}_{\alpha_1\ldots\alpha_n}$:
\be
A^{\mu(-)}_{\alpha_1\ldots\alpha_n}&=&
A^{h=1/2}_{\mu;\alpha_1\ldots\alpha_n}
\!-\!A^{h=3/2}_{\mu;\alpha_1\ldots\alpha_n}\;,
\nonumber\\
A^{h=1/2}_{\mu;\alpha_1\ldots\alpha_n}&=&
V^{(1-)\mu}_{\alpha_1\ldots\alpha_n}
\big (g_1(s)-R\,g_3(s)\big )\;,
\nonumber \\
A^{h=3/2}_{\mu;\alpha_1\ldots\alpha_n}&=&
-\left (V^{(3-)\mu}_{\alpha_1\ldots\alpha_n}\!+\!R
V^{(1-)\mu}_{\alpha_1\ldots\alpha_n}\right )
g_3(s)\;,
\ee
where the factor $R$ is given by
\be
R=-\frac{1}{2\vec k^2}\frac{\alpha_{n-1}}{\alpha_{n+1}}=
  -\frac{1}{2\vec k^2}\frac{n(n+1)}{(2n-1)(2n+1)}\,.
\ee
Here again, the signs for the helicity $1/2$ amplitudes are taken to
correspond to the multipole definition. The widths defined by the helicity
amplitudes are equal to
\be
M\Gamma^{\frac 12}&\!=\!&\rho(s)W^-_{1,1}
\!\Big (g_1(s)-R\,g_3(s)\!\Big )^2\;,
\nonumber \\
M\Gamma^{\frac 32}&\!=\!&\rho(s)\!
\Big(W^-_{3,3}\!+\!R\, W^-_{1,3} \Big )
g_3^2(s)\;,
\ee
and therefore
\be
|A_n^{\frac 12}|^2&\!=\!&\alpha_{n+1}
\rho(s)
\frac{\pi\chi}{m^2_N} |\vec k|^{2n}
\Big (g_1(s)-R\,g_3(s)\Big )^2\;,
\nonumber \\
|A_n^{\frac 32}|^2&\!=\!&\alpha_{n-1}
\frac{(n\!+\!1)(n\!+\!2)}{4(4n^2\!-\!1)}
\rho(s)
\frac{\pi\chi}{m^2_N}
|\vec k|^{2n-4}g_3^2(s)\;,
\ee
with $\chi=m_N+k_{N0}$.

The vertices with couplings $g_1(s)$ and $g_3(s)$ are formed by
different orbital momenta. For the state with total spin $J$
($n\!=\!J\!-\! 1/2$), the orbital momentum is equal to $L\!=\!n\!+\!1$
for the first decay ($s\!=\!1/2$) and $L\!=\!n\!-\!1$ for
the second one ($s=3/2$). Using the Blatt-Weisskopf form factors for the
normalization, we obtain
\be g_1(s)=\frac{g_{1/2}}{F(n\!+\!1,|\vec
k|^2,r)},\quad
g_3(s)=\frac{g_{3/2}}{F(n\!-\!1,|\vec k|^2,r)}\,.~~
\ee

\subsection{Single meson photoproduction}

General structure of the single--meson photoproduction
amplitude in c.m.s. of the reaction is given by
\be
J_\mu\!=\!
i {\mathcal F_1}
 \sigma_\mu\! +&&\!\!{\mathcal F_2} (\vec \sigma \vec q)
\frac{\varepsilon_{\mu i j} \sigma_i k_j}{|\vec k| |\vec q|}
\!+\!i {\mathcal F_3} \frac{(\vec \sigma \vec k)}{|\vec k| |\vec q|}
q_\mu \!+\!i {\mathcal F_4} \frac{(\vec \sigma \vec q)}{\vec q^2}
q_\mu\,, \nn
&&A=\omega^*J_\mu\varepsilon_\mu \omega'\,,
\label{mult_1}
\ee
where $\vec q$ is the momentum of the nucleon in the
$\pi N$ channel and $\vec k$ the momentum of the nucleon in the
$\gamma N$ channel calculated in  the c.m.s. of the reaction. The
$\sigma_i$ are Pauli matrices.

The functions ${\mathcal F_i}$ have the following angular dependence:
\be
{\mathcal F_1}(z) &=
\sum\limits^{\infty}_{L=0}& [LM^+_L\!+\!E^+_L]
P^{\prime}_{L+1}(z)\! +\!
\nn
&&[(L\!+\!1)M^-_L\! +\! E^-_L]
P^{\prime}_{L-1}(z),
\nn
{\mathcal F_2}(z) &= \sum\limits^{\infty}_{L=1}& [(L+1)M^+_L+LM^-_L]
P^{\prime}_{L}(z)   \;,
\nn
{\mathcal F_3}(z) &= \sum\limits^{\infty}_{L=1}& [E^+_L-M^+_L]
P^{\prime\prime}_{L+1}(z) + [E^-_L + M^-_L]
P^{\prime\prime}_{L-1}(z)\;,
\nn
{\mathcal F_4} (z) &= \sum\limits^{\infty}_{L=2}& [M^+_L - E^+_L - M^-_L
-E^-_L] P^{\prime\prime}_{L}(z).
\label{mult_2}
\ee
Here $L$ corresponds to the orbital angular momentum in the
$\pi N$ system,
$P_L(z)$, $P'_L(z)$, $P''_L(z)$ are Legendre polynomials and thier
derivatives,
$z=(\vec k\vec q)/(|\vec  k||\vec q|)$, and $E^\pm_L$ and $M^\pm_L$ are
electric and magnetic multipoles describing transitions to
states with $J=L\pm 1/2$.

The single--meson production amplitude via
the intermediate resonance with $J\!=\!n\!+\!1/2$
(we take pion photoproduction as an example)
has the general form:
\be
A^{i\pm}&&=g_{\pi N}(s)\bar u(q_N)
\tilde N^\pm_{\alpha_1\ldots\alpha_n}(q^\perp)\times
\nn
&&\frac{F^{\alpha_1\ldots\alpha_n}_{\beta_1\ldots\beta_n}}
{M^2-s-iM\Gamma_{tot}}
V^{(i\pm)\mu}_{\beta_1\ldots\beta_n}(k^\perp)
u(k_N) g_i(s) \varepsilon_\mu \;.~~
\label{smp_amp}
\ee
Here, $q_N$ and $k_N$ are the momenta of the nucleon in the $\pi N$
and $\gamma N$ channel and $q^\perp$ and $k^\perp$ are
the components of relative momenta which are orthogonal to
the total momentum of the resonance. The index $i$ lists the
$\gamma N$ vertices given in (\ref{vf_plus}), (\ref{vf_minus}).

The multipole decomposition of the amplitude is given in detail in
\cite{operators}. Below we give the final expressions for
\be
E^\pm_L= E^{\pm(\frac12)}_L + E^{\pm(\frac32)}_L\quad
M^\pm_L= M^{\pm(\frac12)}_L + M^{\pm(\frac32)}_L\,.~~~
\ee
For the '+' states (where $L\!=\!n$):
\be
E^{+(\frac12)}_L&=&
\frac{\sqrt{\chi_i\chi_f}\,\alpha_L}{(2L\!+\!1)(L\!+\!1)}
\frac{g_{\pi N}(s)(|\vec k||\vec q|)^L\,g_1(s)}
{M^2-s-iM\Gamma_{tot}}\,,
\nonumber \\
M^{+(\frac12)}_L&=&E^{+(\frac12)}_L \;,
\nn
E^{+(\frac32)}_L&=&
\frac{\sqrt{\chi_i\chi_f}\;\alpha_L}{(2L\!+\!1)(L\!+\!1)}
\frac{g_{\pi N}(s)(|\vec k||\vec q|)^L\,g_3(s)}
{M^2-s-iM\Gamma_{tot}}\;,
\nonumber \\
M^{+(\frac32)}_L&=&-\frac{E^{+(\frac32)}_L}{L} \;.
\label{mpd_p2}
\ee
Remember that $\chi_i\!=\!m_N\!+\!q_{N0}$ and $\chi_f\!=\!m_N\!+\!k_{N0}$.
For the '-' states, where $L\!=\!n\!+\!1$, the corresponding
equations are
\be
E_L^{-(\frac12)}&=&-
\sqrt{\chi_i\chi_f}\;
\frac{\alpha_L}{L^2}
\frac{g_{\pi N}(|\vec k||\vec q|)^L g_1(s)}
{M^2-s-iM\Gamma_{tot}}\;,
\nn
M_L^{-(\frac12)}&=&-E_L^{-(\frac12)}\;,
\nn
E_L^{-(\frac32)}&=&-\frac{\alpha_{L-2}}{(L\!-\!1)L}
\sqrt{\chi_i\chi_f}\;
\frac{g_{\pi N}|\vec k|^{L-2}|\vec q|^L g_3(s)}
{M^2-s-iM\Gamma_{tot}}\;,
\nn
M_L^{-(\frac32)}&=&0  \;.
\label{mpd_n2}
\ee
These formulae are different from the correspondent expressions given
in \cite{operators} by the factor $(-1)^n$ which enters
now in the resonance propagator. All other formulae
given in \cite{operators} for the single meson photoproduction are not
changed due to this redefinition.

%=============================================================================================

\section{Three-body partial widths of the baryon resonances}

\subsection{Three-body final states}

The total width of the state is calculated by averaging
over polarizations of the resonance and summing over polarizations of
the final--state particles. For the three--particle final
state, the amplitude squared depends on three invariants, total
momentum squared and two intermediate momenta squared $s_{ij} =(q_i +
q_j)^2$. After performing the integration over other variables we
obtain
\be \frac{(2\pi)^4}{2}d\Phi_3(P,q_1,q_2,q_3)\!= \!\int\!
\frac{1}{32s(2\pi)^3} ds_{12}ds_{23}\, .~~ \label{d_g}
\ee
The integration limits
for $s_{12}$ are $(m_1+m_2)^2$ and $(\sqrt s -m_3)^2$.
For given value of $s_{12}$, the limitsfor $s_{23}$ are defined
as follows:
\be
s^{\pm}_{23}\!=\!(E_2\!+\!E_3)^2\!
-\!\Big(\sqrt{E_2^2\! -\!m_2^2} \pm \sqrt{E_3^2\! -\!m_3^2}\Big)^2\;,
\nn
E_2 \!=\! \frac{s_{12}\!-\!m_1^2+\! m_2^2}{2\sqrt{s_{12}}}\;, \qquad
E_3 \!=\! \frac{s\!-\!s_{12}\!-\!m_3^2}{2\sqrt{s_{12}}}\;.
\ee
The three body phase space can also be written as a product of the two
two--body phase spaces, e.g. the phase space of particles 1,2 and the
phase space of the (1,2) system and the third particle:
\be
&&\frac{(2\pi)^4}{2}d\Phi(P,q_1,q_2,q_3)=
\frac{(2\pi)^4}{2}d\Phi(q_1\!+\!q_2,q_1,q_2)\times
\nn
&&\frac{(2\pi)^4}{2}d\Phi(P,q_1\!+\!q_2,q_3)\frac{ds_{12}}{\pi}\,.
\ee
This expression is very useful for the cascade decays when a
resonance is accompanied by a 'spectator' particle and then decays into
two particles. Let us write the explicit form of the expression
$Q\otimes Q$ for the width of baryon with spin $J$
($n=J\!-\!1/2$) which decays into a nucleon with momentum $q_3\equiv
q_N$ and a meson resonance which decays subsequently into two mesons
with momenta $q_1$ and $q_2$ (we denote them as $\pi$ and $\eta$).
If the spin of the intermediate resonance is $J_{12}$ ($J_{12}\!=m\!$),
its decay into two pseudoscalar mesons is described by the orbital
momentum $X^{(m)}$, and therefore
\be &&Q_{\mu_1\ldots\mu_n}\otimes
Q_{\nu_1\ldots\nu_n}= \tilde
P_{\mu_1\ldots\mu_n}^{\alpha_1\ldots\alpha_m} \frac{m_N+\hat q_N}{2m_N} \times
\nn
&&\frac{f^{\alpha_1\ldots\alpha_m}_{\beta_1\ldots\beta_m}g(s_{12})}
{M_R^2-s_{12}-iM_R\Gamma^R_{tot}}
X^{(m)}_{\beta_1\ldots\beta_m}(q^\perp _{12})
X^{(m)}_{\xi_1\ldots\xi_m}(q^\perp _{12}) \times
\nn
&&\frac{g(s_{12})f^{\xi_1\ldots\xi_m}_{\eta_1\ldots\eta_m}}
{M_R^2-s_{12}+iM_R\Gamma^R_{tot}}
P_{\eta_1\ldots\eta_n}^{\nu_1\ldots\nu_m}\,.
\ee
Here we denote the mass and the total width of the intermediate
state as $M_R$ and $\Gamma^R_{tot}$, respectively, and its propagator as
$f^{\alpha_1\ldots\alpha_m}_{\beta_1\ldots\beta_m}$ (see
fig.\ref{fig_3body}).
\begin{figure}[ht]
\epsfig{file=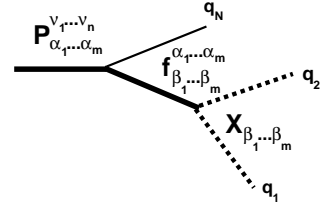,width=11.cm,clip=on}
\vspace{0cm}
\caption{Resonance decay into nucleon and two meson states}
 \label{fig_3body}
\end{figure}
The $g(s_{12})$ is the coupling of the intermediate resonance to the
final--state mesons; the operator $P$ describes the decay of the
initial state into resonance $R$ and spectator nucleon. The operator
$\tilde P$ differs from the operator $P$ by the permutation of
$\gamma$-matrices. The $q^\perp_{12}$ is the relative momentum of the
mesons:
\be &&\tilde g^\perp_{\mu\nu}\!=\!g_{\mu\nu}\!-\! \frac{k_\mu
k_\nu}{s_{12}}\,, \qquad\qquad k=q_1+q_2\,, \nn &&q^\perp_{12\mu}\!=\!
\frac 12 (q_1\!-\!q_2)_\nu \tilde g^\perp_{\mu\nu}\,.
\ee
Using  basic equation (\ref{width}) for the width
\be
f^{\alpha_1\ldots\alpha_m}_{\eta_1\ldots\eta_m}
M_R\Gamma^R_{\pi\eta}=\!\int\!\frac{(2\pi)^4}{2} d\Phi(k,q_1,q_2)
f^{\alpha_1\ldots\alpha_m}_{\beta_1\ldots\beta_m} \times
\nn
g(s_{12})
X_{\beta_1\ldots\beta_m}(q^\perp _{12})
X_{\xi_1\ldots\xi_m}(q^\perp _{12})
g(s_{12})
f^{\xi_1\ldots\xi_m}_{\eta_1\ldots\eta_m}\;,
\ee
we obtain the final expression for the width of the initial resonance
\be
F^{\alpha_1\ldots\alpha_n}_{\beta_1\ldots\beta_n}\,
M\!\Gamma\!=
F^{\alpha_1 \ldots\alpha_n}_{\mu_1 \ldots\mu_n}
\!\int\!\frac{ds_{12}}{\pi}
\frac{(2\pi)^4}{2}d\Phi(P,k,q_3)\times
\nn
g^2(s)\tilde P_{\mu_1\ldots\mu_n}^{\xi_1\ldots\xi_m}
\frac{f^{\xi_1\ldots\xi_m}_{\eta_1\ldots\eta_m}M_R\Gamma^R_{\pi\eta}}
{(M_R^2\!-\!s)^2\!+\!(M_R\Gamma^{R}_{tot})^2}
P_{\nu_1\ldots\nu_n}^{\eta_1\ldots\eta_m}\,.
F^{\nu_1 \ldots\nu_n}_{\beta_1 \ldots\beta_n}\,.~~~~~
\label{rho_n}
\ee
In the limit of zero width of the intermediate state we have
\be
\int\!\!\frac{ds_{12}}{\pi}\frac{M_R\Gamma^R_{tot}}
{(M_R^2\!-\!s_{12})^2\!+\!(M_R\Gamma^R_{tot})^2}\!=\!\!\!
\int\!\!ds_{12}\,\delta(M_R^2\!-\!s_{12})~~
\label{delta}
\ee
and equation (\ref{rho_n}) is reduced to the two--body equation
multiplied by the branching ratio of the decay of the intermediate
state, $Br_{\pi\eta}=\Gamma^R_{\pi\eta}/\Gamma^R_{tot}$.

Let us note that provided a resonance has many decay modes (or the mode
can be in different kinematical channels), the decay amplitude can be
written as a vector with components corresponding to these decay
modes. In this case, eq.(\ref{rho_n}) gives us diagonal
transition elements only. To obtain nondiagonal elements
between different kinematical channels, it is necessary
to use general expression for the phase volume (\ref{d_g}).

\begin{figure}[ht]
\centerline{
\hspace{3cm}
\includegraphics[width=70mm]{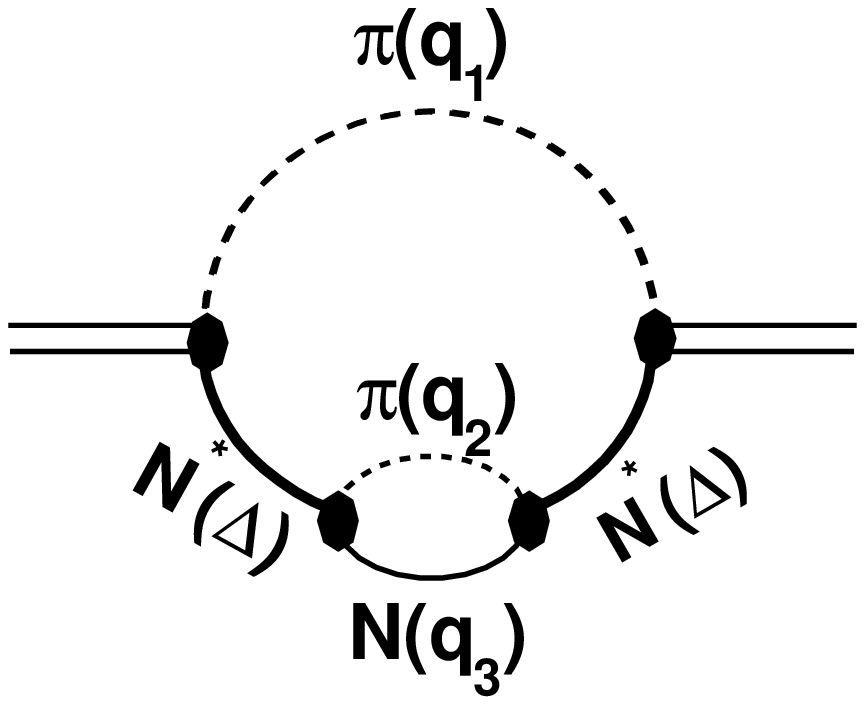} \hspace{-3cm}
\includegraphics[width=70mm]{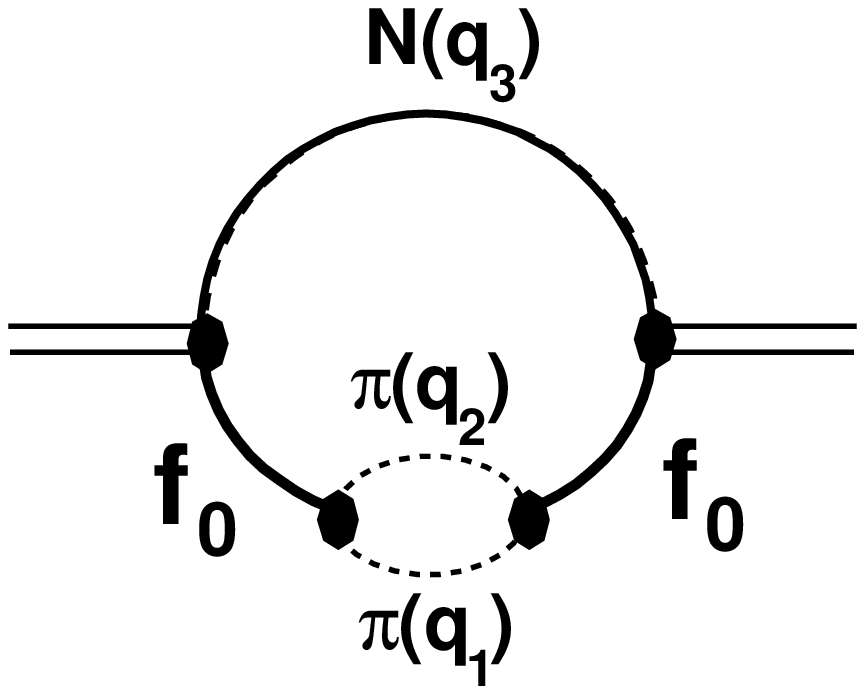}}
\vspace{-1cm}
\caption{$\pi N^*$($\pi \Delta$) and $f_0 N$ loop diagrams}
 \label{loop_3body}
\end{figure}

\subsection{\boldmath $f_0 N$ partial widths of the baryon resonances}

Let us start with the calculation of the $f_0 N$ loop diagram
where $f_0$
denotes a scalar resonance which decays into two pseudoscalar mesons
with momenta $q_1$ and $q_2$ (see fig. \ref{loop_3body}).
The width of the scalar state is defined by
\be
M_R\Gamma^R_{\pi\pi}=g^2_{\pi\pi}(s_{12})\rho(s_{12},m_1,m_2)\,.~~
\ee
Here, $g_{\pi\pi}$ is the coupling of the resonance into two
pseudoscalar mesons. For the simplicity sake,
we denote them as pions but
use different masses $m_1$ and $m_2$. Thus, the final
expressions can be directly used for the decay of any scalar state
into two pseudoscalar particles.

The decay of the baryon state into scalar meson and
nucleon is described by the same vertices as the decay of a baryon into
$\pi N$. The only difference is that due to the positive parity of the
scalar meson, the decay of '+' baryons will be defined by
$N^-_{\mu_1\ldots\mu_n}$ operators and decay of the '-' baryons by the
$N^+_{\mu_1\ldots\mu_n}$ operators:
\be
&P^{(+)}_{\mu_1\ldots\mu_n}&=N^-_{\mu_1\ldots\mu_n}\;,
\nn
&P^{(-)}_{\mu_1\ldots\mu_n}&=N^+_{\mu_1\ldots\mu_n}\;.
\ee
Thus, we can use the results given in
eqs. (\ref{res_piN_plus},\ref{res_piN_minus}) to obtain
\be
M\Gamma^{\pm}_{f_0N}=
\!\!\!\!\!\int\limits_{(m_1+m_2)^2}^{(\sqrt{s}-m_3)^2}
\!\!\!\frac{ds_{12}}{\pi}
\frac{\rho(s,\sqrt{s_{12}},m_N)W^{\pm} M_R\Gamma^R_{\pi\pi}}
{(M_R^2\!-\!s)^2\!+\!(M_R\Gamma^{R}_{tot})^2}\,,~
\ee
where
\be
&W^{(+)} &= \frac {\alpha_{n+1}}{n+1}|\vec q_N|^{2n+2}\;
\frac{m_N\!+\!q_{N0}}{2m_N}g^2_{f_0 N}(s)\;,
\nn
&W^{(-)}& =
\frac {\alpha_n}{2n+1}|\vec q_N|^{2n}\; \frac{m_N\!+\!q_{N0}}{2m_N}
g^2_{f_0 N}(s)\;,
\label{sigmaN}
\ee
and
\be
q_{N0}=\frac{s+m_N^2-s_{12}}{2\sqrt s}\;,\qquad
|\vec q_N|^2=q^2_{N0}\!-\!m_N^2\,.
\label{qn}
\ee

\subsection{\boldmath Vector-meson--$N$ partial widths of baryon
resonances}

The decay vertex of the vector particle into two pseudoscalar particles
is defined by the operator $X^{(1)}_\mu$, and the width is equal to
\be
M_R\Gamma^R_{\pi\pi}=g^2_{\pi\pi}(s_{12})\frac{|\vec q_\pi|^2}{3}
\rho(s_{12},m_1,m_2)\,,~~
\nn
|\vec q_\pi|^2=\frac{
\big (s_{12}\!-\!(m_1\!+\!m_2)^2\big )
\big (s_{12}\!-\!(m_1\!-\!m_2)^2\big )}
{4s_{12}}\;.
\ee
Here, $g_{\pi\pi}$ is the coupling of the vector meson into two
pseudoscalar mesons with masses $m_1$ and $m_2$.

The decay of the baryon into vector meson and nucleon is described by
the same vertices as the decay of baryons into $\gamma N$. In this case,
all three vertices are independent and the width is formed by all
possible transitions between vertices:
\be &&M\Gamma^{\pm}_{\rho N}=
\!\!\!\!\!\int\limits_{(m_1+m_2)^2}^{(\sqrt{s}-m_N)^2}
\!\!\!\frac{ds_{12}}{\pi}
\frac{\rho(s,\sqrt{s_{12}},m_N)W^{\pm} M_R\Gamma^R_{\pi\pi}}
{(M_R^2\!-\!s)^2\!+\!(M_R\Gamma^{R}_{tot})^2}\,,~
\nn
&&W^{\pm}\!=\sum\limits_{i,j=1}^{3} W^{\pm}_{i,j}g_i(s)g^*_j(s)\;,
\ee
where we assume that couplings can be complex magnitudes and
the $W^{\pm}_{i,j}$ functions are defined as follows:
\be
F^{\alpha_1\ldots\alpha_n}_{\beta_1\ldots\beta_n}\,
W^{\pm}_{i,j}\!=
F^{\alpha_1 \ldots\alpha_n}_{\mu_1 \ldots\mu_n}
\!\int\!\frac{d\Omega}{4\pi}
\tilde V^{(i\pm)\xi}_{\mu_1\ldots\mu_n}
\frac{m_N+\hat q_N}{2m_N} \times
\nn
\left (
\frac{k_\xi k_\eta}{s_{12}}-g_{\xi\eta}
\right )
V^{(i\pm)\eta}_{\nu_1\ldots\nu_n}
F^{\nu_1\ldots\nu_n}_{\beta_1 \ldots\beta_n}\,.~~~~
\label{width_rhoN}
\ee
The vertex functions $V^{(i\pm)}$ are given by eqs.
(\ref{vf_plus},\ref{vf_minus}) and $k=q_1+q_2$.

For the '+' states the $W$ functions are given by
\be
W^+_{1,1}&=&\frac{\alpha_{n}}{2n\!+\!1}|\vec q_N|^{2n}
\frac{m_N\!+\!q_{N0}}{2m_N}
\big( 3+\frac{|\vec q_N|^2}{s_{12}}\big )\;,
\nonumber\\
W^+_{1,2}&=&-\frac{\alpha_{n}}{n\!+\!1}
\frac{|\vec q_N|^{2n\!+\!4}}{s_{12}}
\frac{m_N\!+\!q_{N0}}{2m_N}\;,
\nonumber\\
W^+_{1,3}&=&\frac{\alpha_{n}}{2n\!+\!1}|\vec q_N|^{2n}
\frac{m_N\!+\!q_{N0}}{2m_N}
\big( 2+\frac{|\vec q_N|^2}{s_{12}}\big )\;,
\nonumber\\
W^+_{2,2}&=&\frac{\alpha_{n+1}|\vec q_N|^{2n+4}}{n\!+\!1}
\frac{m_N\!+\!q_{N0}}{2m_N}
\big(\frac{2n\!+\!3}{n\!+\!2}+\frac{|\vec q_N|^2}{s_{12}}\big )\;,
\nonumber \\
W^+_{2,3}&=-&\frac{\alpha_{n+1}}{2n\!+\!1}
\frac{|\vec q_N|^{2n+4}}{s_{12}}
\frac{m_N\!+\!q_{N0}}{2m_N}\;,
\nonumber\\
W^+_{3,3}&=&\frac{\alpha_{n}|\vec q_N|^{2n}}{n}
\frac{m_N\!+\!q_{N0}}{2m_N}
\big( 1+\frac{n|\vec q_N|^2}{(2n\!+\!1)s_{12}}\big )\;,
\label{wrho_plus}
\ee
and the '-' states  by
\be
W^-_{1,1}&=&\frac{\alpha_{n+1}}{n\!+\!1}|\vec q_N|^{2n+2}
\frac{m_N\!+\!q_{N0}}{2m_N}
\big( 3+\frac{|\vec q_N|^2}{s_{12}}\big )\;,
\nonumber\\
W^-_{1,2}&=&\frac{|\vec q_N|^{2n+2}}{n\!+\!1}
\frac{m_N\!+\!q_{N0}}{2m_N}
\big( \alpha_{n+1}+\alpha_n\frac{|\vec q_N|^2}{s_{12}}\big )\;,
\nonumber\\
W^-_{1,3}&=&-\frac{\alpha_{n-1}}{n\!+\!1}
\frac{|\vec q_N|^{2n+2}}{s_{12}}\frac{m_N\!+\!q_{N0}}{2m_N}\;,
\nonumber\\
W^-_{2,2}&=&\frac{|\vec q_N|^{2n+2}}{2n\!+\!1}
\frac{m_N\!+\!q_{N0}}{2m_N}
\big(\alpha_{n+1}+\alpha_n\frac{|\vec q_N|^2}{s_{12}}\big )\;,
\nonumber \\
W^-_{2,3}&=&-\frac{\alpha_{n-1}}{2n\!+\!1}
\frac{|\vec q_N|^{2n+2}}{s_{12}}\frac{m_N\!+\!q_{N0}}{2m_N}\;,
\nonumber \\
W^-_{3,3}&=&\frac{\alpha_{n-1}|\vec q_N|^{2n-2}}{2n\!-\!1}
\frac{m_N\!+\!q_{N0}}{2m_N}
\big( 1+\frac{n|\vec q_N|^2}{(2n\!+\!1)s_{12}}\big )\,.~~~
\label{wrho_minus}
\ee
The $q_{N0}$ and $|\vec q_N|$ are defined by eq. (\ref{qn}).

\subsection{\boldmath The $\pi P_{11}$ and $\pi S_{11}$
partial widths of baryon resonances}

Let us calculate the partial width of the baryon resonance decaying
into pseudoscalar meson and $P_{11}$ state (e.g. the Roper
resonance) which in turn decays into pseudoscalar meson and nucleon
with the momenta $q_2$ and $q_3\equiv q_N$ (see fig.\ref{loop_3body}). The
partial width of the $P_{11}$ state is defined by the equation
(\ref{res_piN_minus}) at $n=0$. Therefore,
\be
M\Gamma^{\pm}=
\!\!\!\!\!\int\limits_{(m_2+m_3)^2}^{(\sqrt{s}-m_1)^2}
\!\!\!\frac{ds_{23}}{\pi}
\frac{\rho(s,\sqrt{s_{23}},m_N)W^{\pm} M_R\Gamma^R_{\pi N}}
{(M_R^2\!-\!s_{23})^2\!+\!(M_R\Gamma^{R}_{tot})^2}\,,~
\label{bm_width}
\ee
where
\be
M_R\Gamma^R_{\pi N} =|\vec q_N|^{2}\; \frac{m_N\!+\!q_{N0}}
{2m_N} \rho(s_{23},m_2,m_N)g_{\pi N}^2(s_{23})\;,
\nn
q_{N0}=\frac{s_{23}+m_N^2-m_2^2}{2\sqrt s_{23}}\;,\;\;
|\vec q_N|^2=q^2_{N0}\!-\!m_N^2\;,~~~~
\ee
and $W^\pm$ are defined by
eqs. (\ref{res_piN_plus},\ref{res_piN_minus}). The mass and
momentum of the nucleon should be substituted by the running mass
$\sqrt{s_{23}}$ and
momentum of the $P_{11}$ state $k$:
\be
&W^+ &= \frac {\alpha_n}{2n+1}|\vec k|^{2n}\; \frac{k_{0} +
\sqrt{s_{23}}}{2\sqrt{s_{23}}} g^2(s)\;,
\nn
&W^-& = \frac {\alpha_{n+1}}{n+1}|\vec k|^{2n+2}\;
\frac{k_{0} + \sqrt{s_{23}}}{2\sqrt{s_{23}}}
g^2(s)\,,
\label{pi_P11}
\ee
with
\be
k_{0}=\frac{s+s_{23}-m_{1}}{2\sqrt s}\;,\qquad
|\vec k|^2=k^2_{0}\!-\!s_{23}\,.
\ee
The partial width of the $S_{11}$ state into $\pi N$ is defined by
eq. (\ref{res_piN_plus}) with $n=0$:
\be
M_R\Gamma^R_{\pi N} =\frac{m_N\!+\!q_{N0}}
{2m_N} \rho(s_{23},m_2,m_N)g_{\pi N}^2(s_{23})\;.
\ee
The baryon decay into $\pi S_{11}$ differs from the decay
into $\pi P_{11}$ by P--parity and can be calculated by substituting
$N^{\pm}$ operators by $N^{\mp}$ operators. Thus,
\be
&W^+& = \frac {\alpha_{n+1}}{n+1}
|\vec k|^{2n+2}\; \frac{k_{0} + \sqrt{s_{23}}}{2\sqrt{s_{23}}}
g^2(s)\;,
\nn
&W^- &= \frac {\alpha_n}{2n+1}|\vec k|^{2n}\; \frac{k_{0} +
\sqrt{s_{23}}}{2\sqrt{s_{23}}} g^2(s)\,.
\label{pi_S11}
\ee

%=============================================================================================

\subsection{\boldmath The $\pi \Delta(3/2^+)$ partial widths of
the baryon resonances}

The baryon partial width into $\Delta\pi$, where the
$\Delta$ resonance is produced in the channel (23), can be written as
follows: \be M\Gamma^{\pm}=
\!\!\!\!\!\int\limits_{(m_2+m_3)^2}^{(\sqrt{s}-m_1)^2}
\!\!\!\frac{ds_{23}}{\pi}
\frac{\rho(s,\sqrt{s_{23}},m_N)W^{\pm} M_R\Gamma^R_{\pi N}}
{(M_R^2\!-\!s_{23})^2\!+\!(M_R\Gamma^{R}_{tot})^2}\,.~
\ee
The partial width of the $\frac 32^+$ state is defined by
eq.(\ref{res_piN_plus}) with n=1:
\be
M_R\Gamma^R_{\pi N}\! =\!\frac{|\vec q_N|^2}{3}\; \frac{m_N\!+\!q_{N0}}
{2m_N} \rho(s_{23},m_2,m_3)g_{\pi N}^2(s_{23})\,,~~~
\nn
q_{N0}=\frac{s_{23}+m_N^2-m_2^2}{2\sqrt s_{23}}\qquad
|\vec q_N|^2=q^2_{N0}\!-\!m_N^2\,.~~~~~~
\ee

The decay of baryons with the total spin $J\ge 3/2$ into
pseudoscalar meson and baryon with spin $3/2$ has two vertices which
are defined by different combinations of spins in the intermediate
state and the orbital momentum between this state and a spectator
particle. The decay of '+' states is now described by the
follows operators
\be
&&P^{(1+)\mu}_{\alpha_1\ldots\alpha_n}\Psi_\mu= i\gamma_5 \gamma_\nu
X^{(n+2)}_{\mu\nu\alpha_1\ldots\alpha_{n}}\Psi_\mu   \;,
\nn
&&P^{(2+)\mu}_{\alpha_1\ldots\alpha_n}\Psi_\mu=
i\gamma_5 \gamma_\nu
X^{(n)}_{\nu \alpha_2\ldots\alpha_n}g^\perp_{\alpha_1 \mu}
\Psi_{\mu}  \;.
\label{vf_pD_plus}
\ee

\begin{figure}[ht]
\vspace{-0.5cm}
\epsfig{file=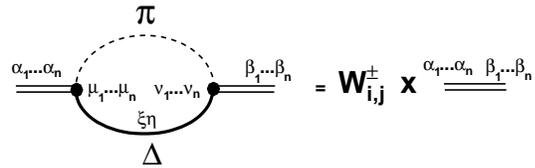,width=8.cm,clip=on}
\vspace{-1cm}
\caption{Index definition in the $\pi \Delta$ loop diagram}
 \label{loop_piDelta}
\end{figure}

The operators for the decay of the '-' states are given by
\be
&&P^{(1-)\mu}_{\alpha_1\ldots\alpha_{n}}\Psi_\mu=
X^{(n+1)}_{\mu\alpha_1\ldots\alpha_{n}}\Psi_\mu \;,
\nn
&&P^{(2-)\mu}_{\alpha_1\ldots\alpha_{n}}\Psi_\mu=
X^{(n-1)}_{\alpha_2\ldots\alpha_{n}} g^\perp_{\alpha_1 \mu}\Psi_\mu \;.
\label{vf_pD_minus}
\ee
The $W^{\pm}$ functions are defined by the squared sum of the decay
amplitudes
\be
W^{\pm}\!=\sum\limits_{i,j=1}^{2} W^{\pm}_{i,j}g_i(s)g^*_j(s)\,,~~
\ee
where $W^{\pm}_{i,j}$ are defined as follows
(see fig.\ref{loop_piDelta} for the
index definition):
\be
F^{\alpha_1\ldots\alpha_n}_{\beta_1\ldots\beta_n}\,
W^{\pm}_{i,j}\!=
F^{\alpha_1 \ldots\alpha_n}_{\mu_1 \ldots\mu_n}
\!\int\!\frac{d\Omega}{4\pi}
\tilde P^{(i\pm)\xi}_{\mu_1\ldots\mu_n} \times
\nn
\left (
\frac{\gamma^\perp_\xi\gamma^\perp_\eta}{3}\!-\!
\tilde g^\perp_{\xi\eta}
\right )
\frac{\sqrt{s_{23}}+\hat k}{2 \sqrt{s_{23}}}
P^{(i\pm)\eta}_{\nu_1\ldots\nu_n}
F^{\nu_1\ldots\nu_n}_{\beta_1 \ldots\beta_n}\,.~~~~
\label{width_dpi}
\ee
Here, $k=q_2+q_3$, and the metric tensor orthogonal
to the momentum of $\Delta$ is equal to
\be
\tilde g^\perp_{\xi\eta}=
g_{\xi\eta}\!-\!\frac{k_\xi k_\eta}{s_{23}}\;, \qquad
\gamma^\perp_\mu=\gamma_\nu\,\tilde g^\perp_{\mu\nu}\,.~~~
\ee
It is useful to extract from the numerator of the
$\Delta$ propagator the nonorthogonal part
\be
\frac{\gamma_\xi\gamma_\eta}{3}-
g_{\xi\eta}\,.~~
\label{nonorth}
\ee
Then, the equation for $W^{\pm}_{i,j}$ can be rewritten as follows:
\be
&&W^{\pm}_{i,j}\;=\;Z^{(1\pm)}_{i,j} + \frac{|\vec
k|^2}{3s_{23}} Z^{(2\pm)}_{i,j}\;,
\nn
k_{0}=&&\frac{s+s_{23}-m_{1}}{2\sqrt s}\;,\qquad
|\vec k|^2=k^2_{0}\!-\!s_{23}\;,
\ee
where $Z^{(1)}_{i,j}$ is defined as a contribution from
(\ref{nonorth}), and the rest is the difference between a full
expression and eq. (\ref{nonorth}). The second part should be
proportional to the momentum of $\Delta$ which we extracted in
explicit form. In the decay of the baryon with the mass not far from
the $\Delta\pi$ threshold
 the contribution
of this part should be much smaller than
the contribution of the nonorthogonal part.

After some cumbersome calculations (an example is given in
Appendix), we obtain for the '+' states
\be
Z^{(1+)}_{1,1}
&=& |\vec k|^{2(n+2)}
\frac{\alpha_{n+2}}{n+1}
\frac{\sqrt{s_{23}} + k_0}{2\sqrt{s_{23}}}\;,
\nn
Z^{(2+)}_{1,1} &=&|\vec k|^{2(n+2)}
\frac{\alpha_{n+1}}{n+1}
\frac{2\sqrt{s_{23}} + k_{0}}{\sqrt{s_{23}}}\;,
\nn
Z^{(1+)}_{2,2} &=& |\vec k|^{2n}
\frac{\alpha_{n}(2n+3)}{3n(2n+1)}
\frac{\sqrt{s_{23}} + k_{0}}{2 \sqrt{s_{23}}}\;,
\nn
Z^{(2+)}_{2,2} &=& |\vec k|^{2n}\frac{\alpha_{n}}{2n+1}
\frac{k_{0}}{\sqrt{s_{23}}}\;,
\nn
Z^{(1+)}_{1,2} &=& 0\;,
\nn
Z^{(2+)}_{1,2} &=& -|\vec k|^{2(n+1)}
\frac{\alpha_{n+1}}{2n+1}
\frac{\sqrt{s_{23}} + k_{0}}{\sqrt{s_{23}}}\,,
\ee
and for the '-' states
\be
Z^{(1-)}_{1,1} &=& |\vec k|^{2(n+1)}
\frac{\alpha_{n+1}(n+2)}{3(n+1)(2n+1)}
\frac{\sqrt{s_{23}} + k_{0}}{2\sqrt{s_{23}}}\;,
\nn
Z^{(2-)}_{1,1} &=& |\vec k|^{2(n+1)}
\frac{\alpha_{n}}{2n+1}
\frac{\sqrt{s_{23}} + k_{0}(n\!+\!1)}{\sqrt{s_{23}}(n\!+\!1)}\;,
\nn
Z^{(1-)}_{2,2} &=&
|\vec k|^{2(n-1)}
\frac{\alpha_{n-1}}{2n-1}
\frac{\sqrt{s_{23}} + k_{0}}{2 \sqrt{s_{23}}}\;,
\nn
Z^{(2-)}_{2,2} &=& |\vec k|^{2(n-1)}
\frac{\alpha_{n-1}n}{(2n-1)(2n+1)}
\frac{2\sqrt{s_{23}} + k_{0}}{\sqrt{s_{23}}}\;,
\nn
Z^{(1-)}_{1,2} &=& 0\;,
\nonumber\\
Z^{(2-)}_{1,2} &=&-|\vec k|^{2n}\frac{\alpha_{n-1}}{2n+1}
\left(\frac{\sqrt{s_{23}} + k_{0}}{\sqrt{s_{23}}} +\frac{1}{2(n\!+\!1)}
\right )\,.
\ee

\subsection{\boldmath The $\pi 3/2^-$ partial widths of
baryon resonances}

In the same way, one can calculate partial widths for the
resonance decay
into $\pi 3/2^-$ states ($\pi D_{13}$ or $\pi D_{33}$).
The partial width of the $\frac 32^-$ state (which belongs to
the '-' states) is defined by eq. (\ref{res_piN_minus}) with n=1:
\be
M_R\Gamma^R_{\pi N}\! =\!
\frac{3|\vec q_N|^4}{4}\; \frac{m_N\!+\!q_{N0}}
{2m_N} \rho(s_{23},m_2,m_3)g_{\pi N}^2(s_{23})\,.~~~
\ee
The decay of the initial--state baryon into $\pi\,3/2^-$ differs
from the $\pi 3/2^+$ decay by the P-parity requiring the substitution of
$W^{\pm}$ by $W^{\mp}$:
\be
W^+_{i,j}(\pi\,3/2^-)=W^-_{i,j}(\pi\,3/2^+)\;,
\nn
W^-_{i,j}(\pi\,3/2^-)=W^+_{i,j}(\pi\,3/2^+).
\ee

%=============================================================================================

\subsection{Singularities in the cascade decay}

The one of the most important point in the partial wave analysis is to
define the position of amplitude singularities in the
complex plane of the energy squared. Resonances correspond to
pole singularities of the transition amplitude, while the
production of
the two or more particles corresponds to threshold singularities. Every
threshold singularity creates two sheets in the complex plane $s$. In
case of the two particle amplitude, these sheets
may be reached by different paths: either directly down to the
complex plane or around threshold singularity.
For the resonance cascade decay into
three particles, the structure is more complicated, and different sheets
may be reached by the choice of the integral path in the analytical
expressions for the resonance partial widths given above.

Consider as an example the structure of the
singularities in the decay of a resonance into the $\pi\Delta(1232)$
channel. The imaginary part of the Breit-Wigner pole is defined by the
discontinuity of the diagram shown in Fig.~\ref{loop_3body}.
\be
\frac{g_{\pi\Delta}^2}{M^2-s-i\,g^2_{\pi\Delta} \rho_3(s)}
\ee
If the intermediate resonance decays into particles 2 and 3 (with
momenta $k_2$ and $k_3$, $s_{23}=(k_2+k_3)^2$), the
three-body phase volume $\rho_3(s)$ is defined as follows:
\be
&&\rho_3(s)=\!\!\int\limits_{(m_2+m_3)^2}^{(\sqrt{s}-m_1)^2}
\!\!\frac{ds_{23}}{\pi} I(s,s_{23})\;,
\nn
&&I(s,s_{23})=\frac{\rho(s,\sqrt{s_{23}},m_1)\,M_R\Gamma^{R}_{tot}}
{(M_R^2\!-\!s_{23})^2\!+\!(M_R\Gamma^{R}_{tot})^2}\;,~~
\nn
&&M_R\Gamma^{R}_{tot}=\rho(s_{23},m_2,m_3)g^2(s_{23})\;,
\label{rho3}
\ee
where the $\rho$-functions are given by
eqs. (\ref{phv_2b0},\ref{phv_2b}).

The function $\rho_3(s)$ has two threshold singularities in the complex
plane of total energy squared $s$. The first threshold singularity
originates at the squared sum of the final-state particle masses, at
\be
s=(m_1+m_2+m_3)^2\,,
\ee
and physically it corresponds to the
possibility of the system to decay into three particles
(three-particle singularity in the amplitude).
Another square root singularity is located in the
complex plane and starts at:
\be s_{cut}&=&a_{cut}-i\,b_{cut}=(
\sqrt{M^2_R-iM_R\Gamma^R_{tot}}+m_1)^2
\nn
&\simeq&(M_R+m_1-i \Gamma_{tot}^R/2)^2\ .
\label{a_cut}
\ee
It corresponds to the decay
of the system into a particle with mass $m_1$ and  resonance
with complex mass $M_R-i\Gamma^R_{tot}/2$ (usually we assume
$\Gamma^R_{tot}<<M_R$).
\begin{figure}[h!]
\centerline{
\epsfig{file=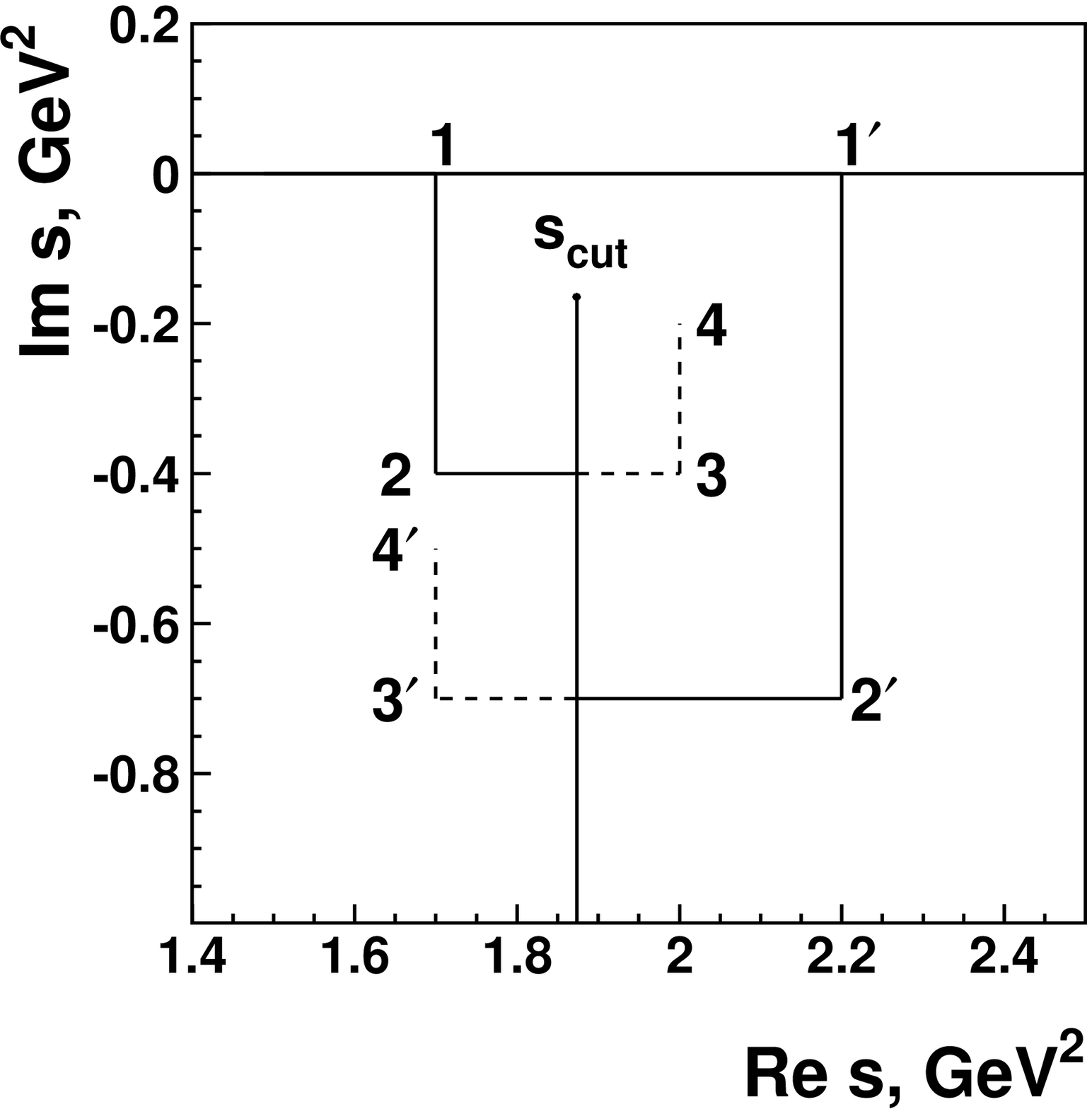,width=0.24\textwidth,clip=on}
\epsfig{file=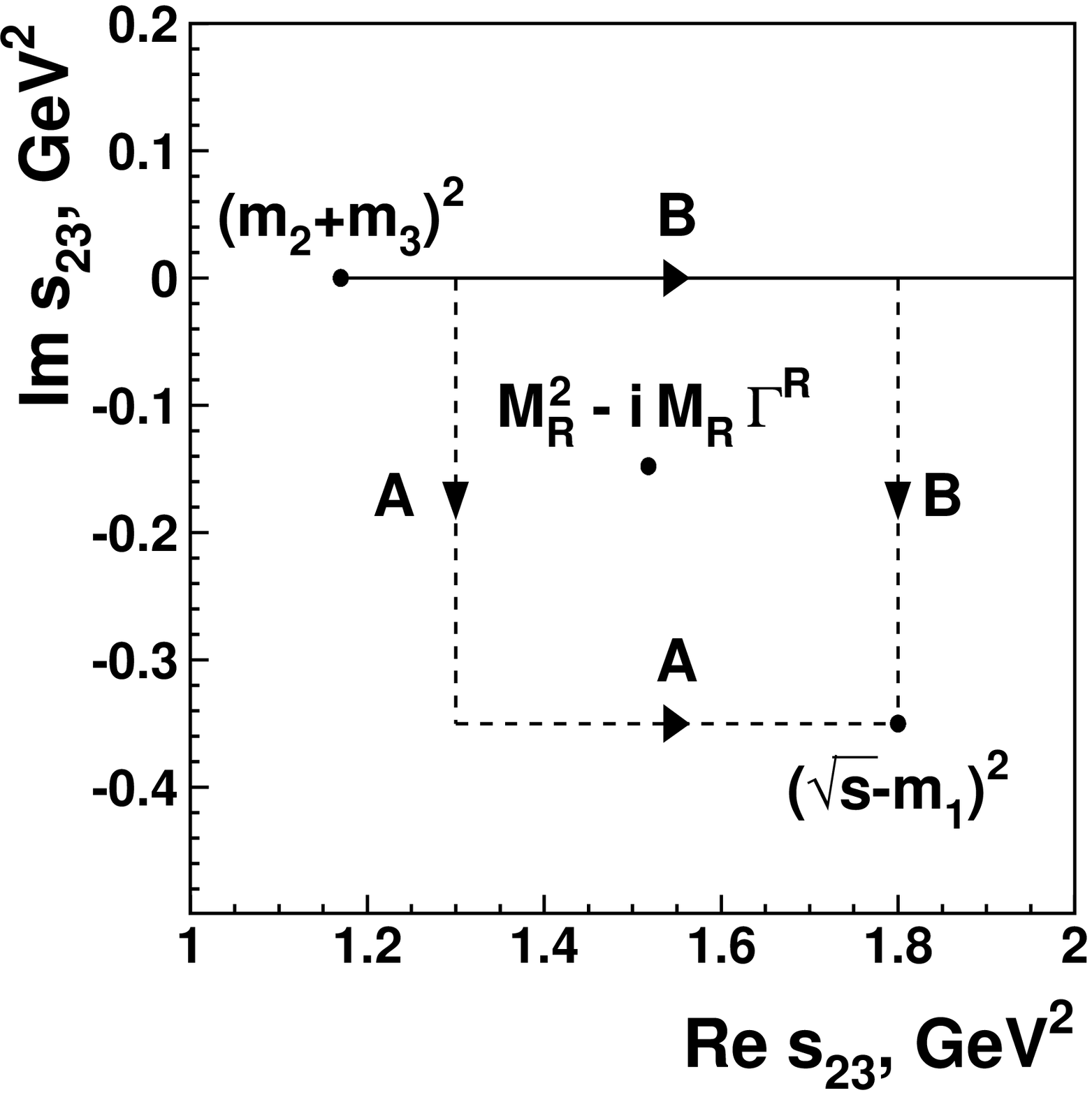,width=0.24\textwidth,clip=on}}
\caption{The structure of threshold
singularities in the $s$ and $s_{23}$ complex planes for the
$\pi\Delta$ channel.}
\label{cut}
\end{figure}

If the width of the intermediate state ($\Delta$) tends to zero
$\Gamma^R_{tot}\!\to\! 0$, the intermediate resonance turns into a stable
particle, and the contribution from the three--particle cut disappears.
For the energy region above the
three--particle threshold but below the two--particle one,
\be
(m_1+m_2+m_3)^2<s<(M_R+m_1)^2\ ,
\ee
the integrand
$I(s,s_{23})$ does not reach the singular point at $s_{23}=M^2_R$, and
it is proportional to $\Gamma^R_{tot}$. In the energy region above the
two--particle threshold, the contribution of $I(s,s_{23})$ is
proportional to the $\delta$-function (see (\ref{delta})). In the
complex plane $s$, the two--body cut moves towards the real axis and
finally becomes the only threshold singularity.

The influence of a pole on  point of the physical region
is defined by the distance between the pole position and this point.
The path used for the calculation of this
distance should not cross any cut. Therefore, if the pole is
located on the sheet closest to the physical region, the minimal
distance is defined by the imaginary part of the pole position.
If the pole is located on another sheet, the minimal distance
to the physical region can be
estimated as a sum of distances ({\it i}) between pole position and the
beginning of the cut and ({\it ii}) between the beginning of the cut and real
axis. Thus, if the pole is located far from the cut, only its position
on the sheet closest to the physical region is important. However, if
the real part of the pole position is close to $a_{cut}$, see
(\ref{a_cut}), the
position on both sheets can be important to explain the behaviour of
the amplitude. For example, in the decay of a resonance into
$\Delta(1232)\pi$ channel, the threshold singularity in the complex
plane starts at $s\sim (1370-i60)^2$ MeV$^2$ (see Fig. \ref{cut}). This
is very close to the position of the Roper resonance. Such a situation
should be handled with care.

Let us formulate some rules which can be used to reach a
desirable sheet in the complex s plane.
The sum of integrals taken between the points 1,2,3,4 reaches the
sheet I, which is the closest one to the physical region below $a_{cut}$
(at $Re\,s<a_{cut}$):
\be
\rho_3(s)\!=\!\!
\int\limits_{(m_2+m_3)^2}^{(\sqrt{s_1}-m_1)^2}
\!\!+\!\!
\int\limits_{(\sqrt{s_1}-m_1)^2}^{(\sqrt{s_2}-m_1)^2}
\!\!+\!\!
\int\limits_{(\sqrt{s_2}-m_1)^2}^{(\sqrt{s_3}-m_1)^2}
\!\!+\!\!
\int\limits_{(\sqrt{s_3}-m_1)^2}^{(\sqrt{s_4}-m_1)^2}
\nn
\frac{ds_{23}}{\pi} I(s,s_{23})\,.~~
\ee
The sum of integrals taken between points $1',2',3',4'$ reaches
the sheet II which is the closest one to the physical region above
$a_{cut}$ (at $Re\,s>a_{cut}$).
In the $s_{23}$ complex plane, the different sheets are reached when the
pole at $M^2_R-iM_R\Gamma_{tot}^R$ is located on different sides of the
integration path (see, for example, the paths (A) and (B) shown on
the right
panel of Fig.~\ref{cut}). The contour integral which can be constructed
from such two paths is equal to the residue of the pole which is
proportional to $\rho(s,M_R-\Gamma^R_{tot}/2,m_1)$. This residue
provides us the difference of the $\rho_3(s)$ function defined on
the two sheets. The sheet closest to the physical region can be
reached by the integration performed first over the real axis and
then by integrating over the imaginary axis in the $s_{23}$ plane. Such an
integration path is close to that given by paths $1,2$ and $1',2'$ in
the $s$-plane and is much more convenient for practical use:
\be
\rho_3(s)&=& \!\!\!\!\int\limits_{(m_2+m_3)^2}^{s_{23}^{int}}
\!\!\!\!\!\frac{ds_{23}}{\pi}
I(s,s_{23})\,+\!\!\!\!\!\!\int\limits_{s_{23}^{int}}^
{(\sqrt{s}-m_1)^2}\!\!\!\!\!\frac{ds_{23}}{\pi}
I(s,s_{23})\;,
\nn
s_{23}^{int}&=&Re\,(\sqrt{s}-m_1)^2.
\label{cont}
\ee
When $\rho_3(s)$ is calculated with this integral path, the function
$Re\,s$ has, at fixed $Im\,s$, the discontinuity
at $Re\,s=a_{cut}$ if $-Im\,s>b_{cut}$.
This is demonstrated in Fig.~\ref{re_im} where
the real (left-hand panel) and imaginary (right-hand panel) parts of
$\rho_3(s)$ are plotted as function of $Re\,s$ at fixed values of
$Im\,s_n$ where $n=0,30$, $Im\,s_n=-20 MeV^2 n$. On the physical axis
($n=0$), the real part of $\rho_3(s)$ is very small up to $a_{cut}$
and the imaginary part vanishes. At $-Im\,s_n<b_{cut}$, $\rho_3(Re\,s)$
is a smooth function while at $-Im\,s_n>b_{cut}$ the function has a
discontinuity at $Re\,s=a_{cut}$ being defined on different sheets.
We plot real and imaginary parts of $\rho_3(Re\,s)$ for every $n$ with
shifting the curves down by the same value $20$ MeV$^2$. It
is seen that when $-Im\,s_n$ reaches $b_{cut}$, the functions have a
well seen discontinuity.

\begin{figure}[h!]
\centerline{
\epsfig{file=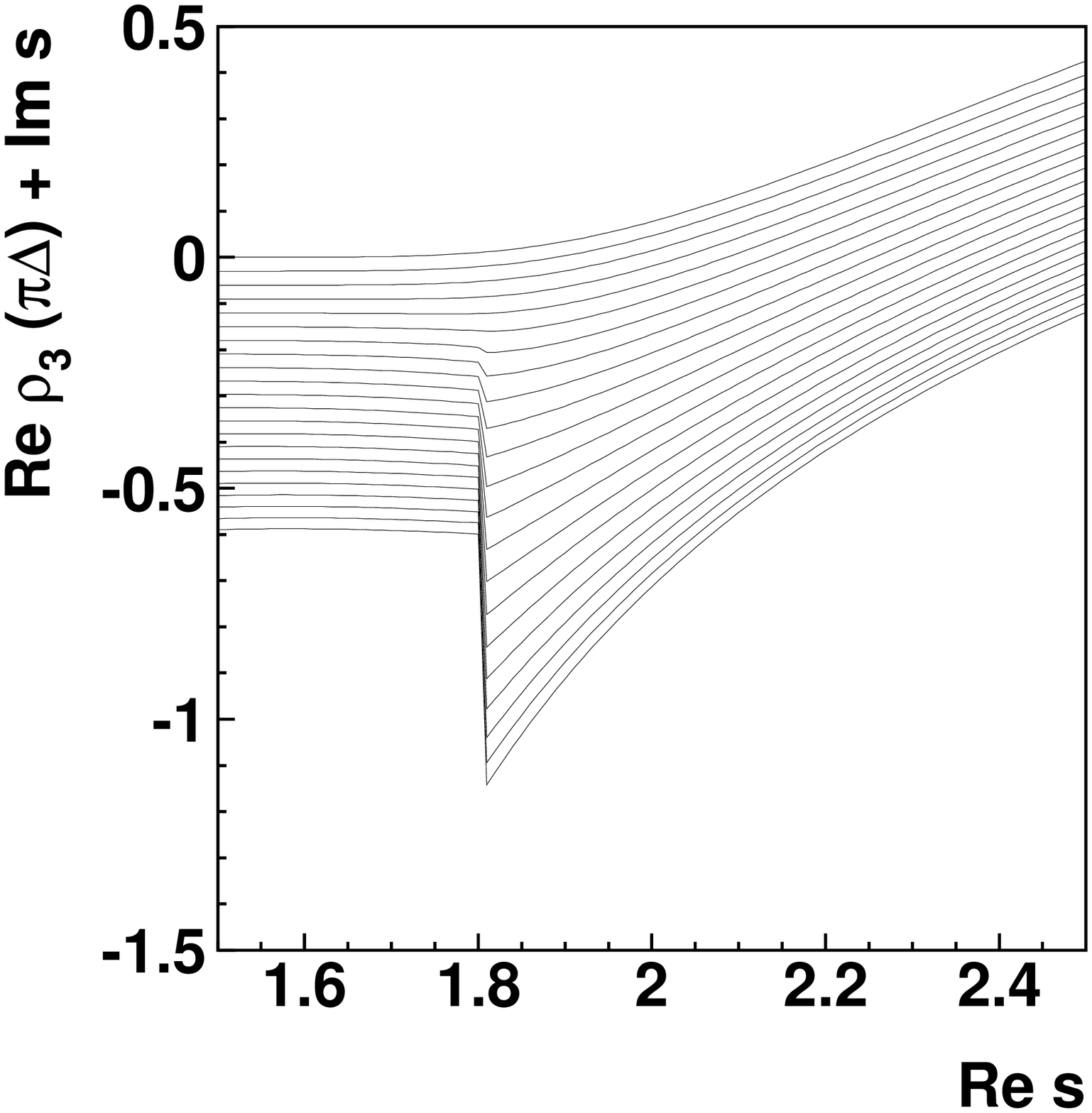,width=0.23\textwidth,clip=on}
\epsfig{file=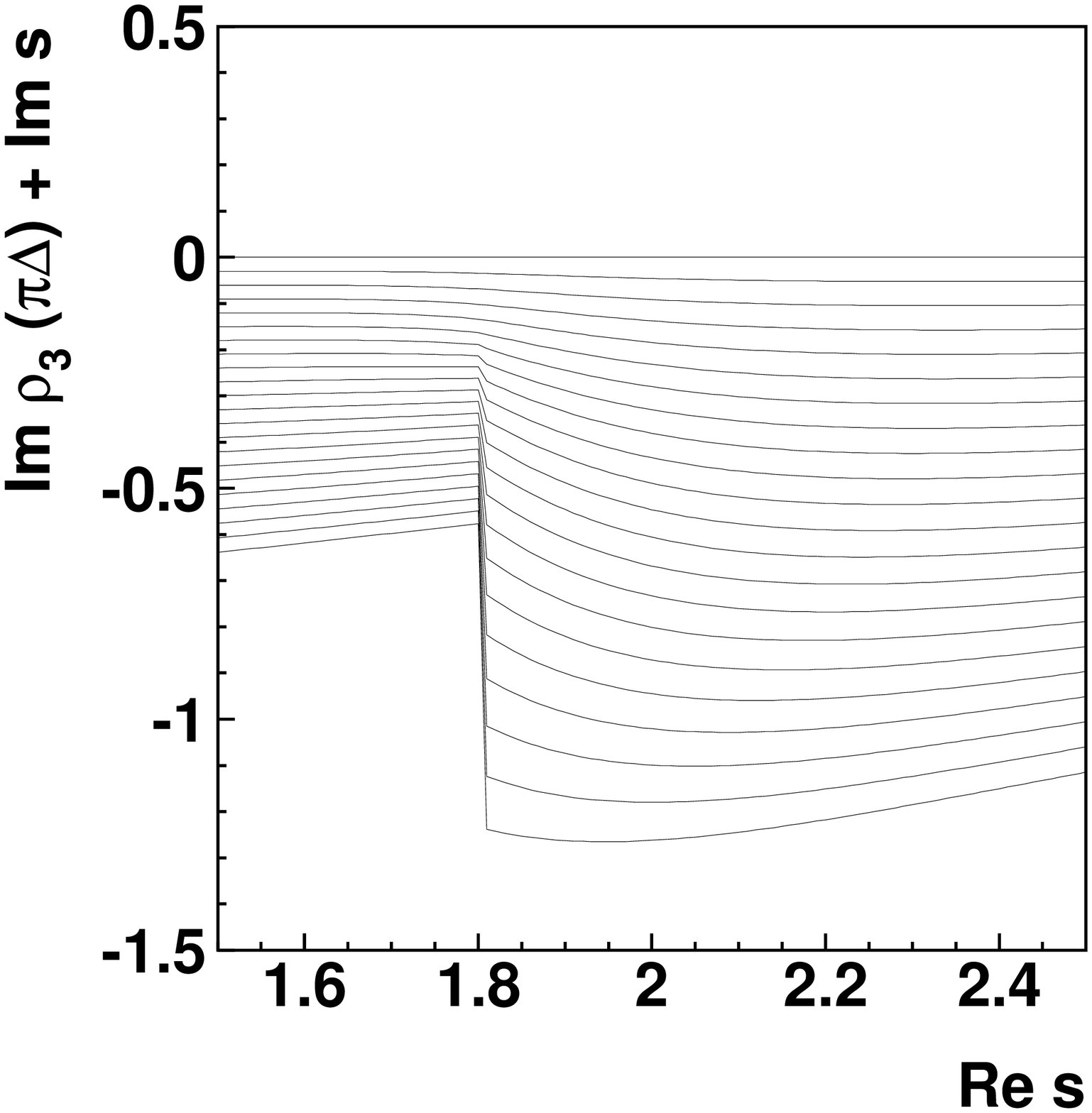,width=0.23\textwidth,clip=on}}
\caption{The real (left-hand panel) and imaginary (right-hand
panel) part of the $\rho_3(s)$ as a function of
$Re\,s$ at fixed $Im\,s_n=-20n\,MeV^2$, $n=1,30$.
For every $n$ the curves are shifted down by the same step
20 MeV$^2$.}
\label{re_im}
\end{figure}

%=============================================================================================

\section{Conclusion}

In this  paper we present further development of
the spin-momentum operator expansion
approach suggested initially in \cite{operators}.
Explicit expressions for cross sections
and resonance partial widths are given for a large number of the
pion-induced and photoproduction reactions with two or three particles
in the final state.  The formulae are given explicitly as they are used
by the Crystal Barrel at ELSA collaboration in the analysis of single
and double meson photoproduction.

\section*{Acknowledgments}
We would like to thank V.V. Anisovich, L.G. Dakhno, E. Klempt and V.A. Nikonov for helpful
discussions and a critical reading of the manuscript. We also thank V.A. Nikonov for the
assistance with numerical calculations and figures.
 The work was supported by the Deutsche
Forschungsgemeinschaft within the Sonderforschungsbereich SFB/TR16.
 We would like to thank the
Alexander von Humboldt foundation for generous support in the initial
phase of the project, A.V.A. for a AvH fellowship and A.V.S. for the
Friedrich-Wilhelm Bessel award. A.~Sarantsev gratefully acknowledges
the support from Russian Science Support Foundation.
This work is also supported by Russian Foundation for Basic Research,
project no 04-02-17091 and RSGSS 5788.2006.2 (Russian State Grant Scientific School).
%=============================================================================
\section*{Appendix}
%=============================================================================

The operator $X^{(n+1)}_{\nu\mu_1\ldots\mu_n}$
can be written as a series of products of metric tensors and
relative momentum vectors. The first term is proportional to the
production of relative momentum vectors $k^\perp_\mu$, other terms
correspond to the substitution of two vectors by a metric
tensor with corresponding indices, and the coefficients are defined to
satisfy the properties of the operator (\ref{x_oper}):
\be
X^{(n+1)}_{\nu\mu_1\ldots\mu_n}(k^\perp) =\alpha_{n+1} \bigg [
k^\perp_{\nu}k^\perp_{\mu_1}k^\perp_{\mu_2}k^\perp_{\mu_3}
\ldots k^\perp_{\mu_n}-
\frac{k^2_\perp}{2n\!+\!1}\times
\nn
\bigg(\sum\limits_{i=1}^n
g^\perp_{\nu\mu_i}\prod\limits_{j\ne i} k^\perp_{\mu_j}+
\sum\limits_{i<j}^n
g^\perp_{\mu_i\mu_j}k^\perp_{\nu}
\!\!\prod\limits_{m\ne i\ne j}\!\!
k^\perp_{\mu_m}+\ldots \bigg )
\nn
+\frac{k^4_\perp}{(2n\!+\!1)(2n\!-\!1)}\bigg(
\sum\limits_{i,j<m}^n
g^\perp_{\nu\mu_i}g^\perp_{\mu_j\mu_m}
\!\!\prod\limits_{l\ne i\ne j\ne m}\!\! k^\perp_{\mu_l}+
\nn
\sum\limits_{i<k,j<m}^n
g^\perp_{\mu_i\mu_k}g^\perp_{\mu_j\mu_m} k^\perp_{\nu}
\prod\limits_{^{l\ne i\ne k}_{\ne j\ne m}} k^\perp_{\mu_l}+
\ldots\bigg)+\ldots\bigg ].~~~~~~~~
\label{x-direct}
\ee
Taking into account the tracelessness and orthogonality
of the fermion propagator (\ref{F_proper}) to $\gamma^\mu$
one obtains:
\be
F^{\mu_1\ldots\mu_n}_{\alpha_1\ldots \alpha_n} \gamma_\nu
X^{(n+1)}_{\nu\alpha_1\ldots\alpha_{n}}\!=\!
F^{\mu_1\ldots\mu_n}_{\alpha_1\ldots \alpha_n}
&\alpha_{n+1}& \hat k^\perp
k^\perp_{\alpha_1}\ldots k^\perp_{\alpha_{n}}\;.~~~~~
\ee
Therefore, for the $\pi N$ widths of the '-' states:
\be
&&F^{\alpha_1\ldots\alpha_n}_{\xi_1\ldots \xi_n}
\int\frac{d\Omega}{4\pi}
i\gamma_\nu \gamma_5 X^{(n+1)}_{\nu\xi_1\ldots\xi_n}(k^\perp)
\frac{\hat k_{N}\!+\!m_N}{2m_N}
i\gamma_5 \gamma_\eta \times
\nn
&&X^{(n+1)}_{\eta\beta_1\ldots\beta_n}(k^\perp)
 F^{\beta_1\ldots\beta_n}_{\nu_1\ldots \nu_n}=-
F^{\alpha_1\ldots\alpha_n}_{\xi_1\ldots \xi_n}\alpha^2_{n+1}
\frac{k_{0N}\!+\!m_N}{2m_N}\times
\nn
&&\int\frac{d\Omega}{4\pi}\hat k^\perp \hat k^\perp
k^\perp_{\xi_1}\ldots k^\perp_{\xi_n}
k^\perp_{\beta_1}\ldots k^\perp_{\beta_n}
F^{\beta_1\ldots\beta_n}_{\nu_1\ldots \nu_n}=-
F^{\alpha_1\ldots\alpha_n}_{\xi_1\ldots \xi_n}\times
\nn
&&\frac{\alpha^2_{n+1}}{\alpha^2_n}
\frac{k_{0N}\!+\!m_N}{2m_N} k^2_\perp
\int\frac{d\Omega}{4\pi}
X^{(n)}_{\xi_1\ldots\xi_n}(k^\perp)
X^{(n)}_{\beta_1\ldots\beta_n}(k^\perp)\times
\nn
&&F^{\beta_1\ldots\beta_n}_{\nu_1\ldots \nu_n}=
F^{\alpha_1\ldots\alpha_n}_{\nu_1\ldots \nu_n}
\frac{\alpha_{n+1}}{n\!+\!1}
\frac{k_{0N}\!+\!m_N}{2m_N} |\vec k|^{2n+2}\,.~~
\ee
Here we used property (\ref{F_proper}) as well as the fact
that integral over even numbers of
$k_\mu^\perp$ vectors is equal to zero.

The convolution of two gamma matrices with orbital operator is
equal to zero due to tracelessness and symmetry properties:
\be
\gamma_{\mu_1} \gamma_{\mu_2} X^{(n)}_{\mu_1\ldots\mu_n}\! =\!
g_{\mu_1\mu_2} X^{(n)}_{\mu_1\ldots\mu_n}\!\!+\!
\sigma_{\mu_1\mu_2} X^{(n)}_{\mu_1\ldots\mu_n}\!=\! 0\,.~~
\ee
Here, the second term is equal to zero being the product of the
antisymmetrical and symmetrical tensors.

Other useful expressions are as follows:
\be
F_{\mu_1\ldots\mu_n}^{\alpha_1\ldots \alpha_n} \gamma_\nu
X^{(n+2)}_{\nu\beta\alpha_1\ldots\alpha_{n}}(k^\perp) &=&
\nonumber\\
F_{\mu_1\ldots\mu_n}^{\alpha_1\ldots \alpha_n}
\alpha_{n+2} &\Big(& \hat k^\perp k^\perp_{\beta}
k^\perp_{\alpha_1}\ldots k^\perp_{\alpha_{n}}-
\nonumber\\
\frac{k_\perp^2}{2n+3}(n g^\perp_{\beta\alpha_1}\hat k^\perp
k^\perp_{\alpha_2}\ldots k^\perp_{\alpha_{n}} &+&
\gamma^\perp_\beta k^\perp_{\alpha_1}\ldots k^\perp_{\alpha_{n}})
\Big)\,,
\ee
\be
F_{\mu_1\ldots\mu_n}^{\nu_1\ldots \nu_n}
\int \frac{d\Omega}{4\pi}
X^{(n+1)}_{\beta\nu_1\ldots\nu_{n}}(k^\perp) &&
X^{(n+1)}_{\beta'\nu_1'\ldots\nu_{n}'}(k^\perp)
F^{\mu_1'\ldots\mu_n'}_{\nu_1'\ldots \nu_n'}\; g_{\beta\beta'}
\nonumber\\
= -F_{\mu_1\ldots\mu_n}^{\mu_1'\ldots \mu_n'}\;|\vec k|^{2n+2}
&&\frac{\alpha_{n+1}}{2n+1}\,,
\ee
\be
F_{\mu_1\ldots\mu_n}^{\nu_1\ldots \nu_n}
\int \frac{d\Omega}{4\pi}
X^{(n+1)}_{\beta\nu_1\ldots\nu_{n}}(k^\perp) &&
X^{(n+1)}_{\beta'\nu_1'\ldots\nu_{n}'}(k^\perp)
F^{\mu_1'\ldots\mu_n'}_{\nu_1'\ldots \nu_n'}\; k^\perp_{\beta}k^\perp_{\beta'}
\nonumber\\
= F_{\mu_1\ldots\mu_n}^{\mu_1'\ldots \mu_n'}\;|\vec k|^{2n+4}
&&\frac{\alpha_{n}}{2n+1}\,,
\ee
\be
F_{\alpha_1\ldots\alpha_n}^{\mu_1\ldots \mu_n}\;g_{\alpha_1\beta_1}
O_{\beta_2\ldots\beta_n}^{\alpha_2\ldots \alpha_n}
F_{\nu_1\ldots\nu_n}^{\beta_1\ldots \beta_n}\;=\;
(-1)^n
F_{\nu_1\ldots\nu_n}^{\mu_1\ldots \mu_n}\,.
\ee

\end{document}